\documentclass[aps,pra,showpacs,twoside,twocolumn,longbibliography,10pt]{revtex4-1}
\usepackage[colorlinks=true, citecolor=red, urlcolor=blue ]{hyperref}
\usepackage{xcolor}
\usepackage{epsfig,newlfont,amssymb,amsfonts,amsmath,bm,subfigure,palatino,mathtools,amsthm,braket,times,soul,enumitem,color}
\usepackage[normalem]{ulem}
\newcommand{\stkout}[1]{\ifmmode\text{\sout{\ensuremath{#1}}}\else\sout{#1}\fi}
\usepackage[english]{babel}
\usepackage[utf8]{inputenc}
\usepackage{array}
\usepackage{graphics}
\newtheorem{theorem}{Theorem}
\newtheorem{corollary}{Corollary}

\newcommand{\ketbra}[2]{|#1\rangle \langle #2|}
\def\Tr{\text{Tr}}
\def\Cov{\text{Cov}}

\begin{document}

\title{%Genuine multiparty entanglement favors efficient estimation of multiple field strengths
%Advantage of non-local probes in estimating multiple field strengths\\
%Genuinely entangled probes provide advantage in estimating independent local fields
%Greenberger-Horne-Zeilinger state is the best probe\\for multiparameter estimation of  independent local fields 
%Quantum features in multiparameter estimation of independent local fields
Precision in estimating independent local fields:\\ attainable bound and indispensability of genuine multiparty entanglement
}

\author{Aparajita Bhattacharyya and Ujjwal Sen}
\affiliation{Harish-Chandra Research Institute, A CI of Homi Bhabha National Institute, Chhatnag Road, Jhunsi, Prayagraj 211 019, India}

\begin{abstract}
Estimation of local quantum fields is a crucial aspect of quantum metrology applications, and often also forms the test-bed to analyze the utility of quantum resources, like entanglement. 
However, so far, this has been analyzed using the same local field for all the probes, and so, although the encoding process utilizes a local Hamiltonian, there is an inherent ``nonlocality" in the encoding process in the form of a common local field applied on all the probes.  
%It is therefore plausible that entanglement is not necessary in such estimation.
We show that estimation of even independent multiple 
%independent 
field strengths of a local Hamiltonian, i.e., one formed by a sum of single-party terms, necessitates the utility of 
%a highly symmetric genuine multiparty entangled state, viz. the Greenberger-Horne-Zeilinger (GHZ) state, 
genuine multipartite entangled state as the input probe. 
%This
The 
feature 
%however 
depends on the choice of the weight matrix considered, which is full-rank and contains non-vanishing off-diagonal terms.
%, to define a figure of merit in the multiparameter estimation. 
We obtain this result by providing a lower bound on the precision of multiparameter estimation, optimized over input probes, for an arbitrary positive semi-definite weight matrix. We find that the bound can be expressed in terms  of the weight matrix and the eigenvalues of the arbitrary multiparty local  encoding Hamiltonian. We show that there exists a  weight matrix for which this bound is always attainable 
by the Greenberger-Horne-Zeilinger (GHZ) state, chosen in a certain basis. Furthermore, we find the parametric form of the most general optimal state for three parties.
%only when the probe is the GHZ state, up to a relative phase. 
%In particular,  
We also show that no pure product state can achieve the lower bound. 
%Indeed, the gap - in precision - between genuinely multiparty entangled  and product states acting as probes, increases with increasing number of parties.
Finally, for an arbitrary weight matrix and an arbitrary multiparty local  encoding Hamiltonian, we prove that using a probe that is in any mixed state  provides a precision lower than that obtainable using pure states.
%for the GHZ state. 
To emphasize the importance of the weight matrix considered, we also prove that the choice of  identity operator as the same - thereby ignoring the 
``off-diagonal'' 
covariances in the precision matrix - does not require the use of genuine multiparty entanglement in input probes for attaining the best precision, and the optimal probe can be a pure product.
\end{abstract}

% 3306 words. (3750 limit)

\maketitle
\section{Introduction}
The goal of quantum  metrology is to  
estimate one or more 
parameters that are encoded via some physical operation onto a system, and to demonstrate an enhancement in the estimation, fueled by one or more quantum resources. 
Enhancement in  precision of estimated parameters due to  the presence of quantum resources is a fundamental tenet
of quantum-enhanced sensing, and has been widely analyzed both theoretically and experimentally~\cite{Wooters,asymptotic,Braunstein1,Matsumoto,macconerev,expt,rev2014,rev2015,Rafal,ref2,ref6,ref1,ref7,mixed,ref8,ref10,ref3,ref9,review1,ref11,ref4,ref5,funda_9,ref12,ref13,ref14,funda2,funda14,ref15,rev0,dur,disorder,dur2,evenodd,Jayanth}.

% Estimation of the common field strength - common over parties - of an encoding Hamiltonian is widely explored~\cite{common_field6,common_field5,common_field4,common_field3,common_field2,common_field1}.
% %while t
% The simultaneous estimation of 
% non-commuting common 
% fields 
% has also been
% %is less 
% delved into~\cite{mult_field1,mult_field_noise1,mult_field_noise2}.
% %For instance, i
% It is known that while estimating the common field strength of an encoding Hamiltonian whose local field terms are Pauli-\(z\), the GHZ state turns out to be the best probe - providing better scaling of the minimum error than optimal product ones~\cite{maccone}.
% Although in these cases, the encoding process utilizes a local Hamiltonian, there is an inherent ``nonlocality" in the encoding process in the form of a common local field applied on all the probes. It may therefore seem plausible that using independent local fields in the encoding process may lead to the ineffectuality of entanglement in the, now multiparameter, estimation.
% %However, h
% Here we prove that even if we estimate multiple independent field strengths of a local Hamiltonian, 
% %then 
% the optimal probe is 
% \textcolor{red}{necessarily genuinely multipartite entangled, with the GHZ state being a potential candidate.}
%the GHZ, up to a relative phase.

Estimating local quantum fields is an important model for checking the utility of quantum resources in quantum metrology. It is known that such estimation can be qualitatively benefitted by employing entanglement as a resource.
Estimation of the common field strength - common over parties - of an encoding Hamiltonian is widely explored~\cite{common_field6,common_field5,common_field4,common_field3,common_field2,common_field1}.
%while t
The simultaneous estimation of 
non-commuting common 
fields 
has also been
%is less 
delved into~\cite{mult_field1,mult_field_noise1,mult_field_noise2}.
%For instance, i
It is known that while estimating the common field strength of an encoding Hamiltonian whose local field terms are Pauli-\(z\), the GHZ state turns out to be the best probe - providing better scaling of the minimum error than optimal product ones~\cite{maccone}.
%However, until now, this has been checked by using the same local field for all the probes, and 
So although the encoding process uses a local Hamiltonian, there is an inherent “nonlocality” in the encoding process in the form of a common local field applied on all the probes. It is therefore plausible that using independent local fields in the encoding process may lead to the futility of entanglement in the estimation. To the contrary, we show that estimation of multiple independent field strengths of a local Hamiltonian, i.e., one formed by a sum of single-party terms, necessitates genuine multipartite entangled states. 
Experimentally, genuine multipartite entangled states can be prepared in photonic systems~\cite{state1,state2,state3} (taken as a specific example). Moreover, there are experiments in photonic systems which  implement entangled measurements~\cite{meas1,meas2,meas3}. Such measurements can be utilized in photonic systems to implement the optimal  measurements required to saturate the multiparameter precision bound. This can potentially lead to the practical implementation of creating genuine multipartite entangled states and entangled measurements, which can be utilized as input probes and measurements respectively in realistic physical scenarios.
%Experimentally, estimation of simultaneous phases originating from field-like terms is not uncommon, where genuine multipartite entangled states such as GHZ states turn out to be fruitful~\cite{expt,exp_valeri,exp_valeri2,exp_giova}.}

We analyze the situation where multiple (local) field  strengths of a many-body encoding Hamiltonian are simultaneously estimated, with each field strength being independent and, in principle, distinct from the others. 
%%%
This is therefore a  scenario of multiparameter estimation, 
%where  simultaneous estimation of multiple parameters need to be considered, 
and unlike the case of assessing a single parameter  where  variance is a useful quantity, the covariance matrix of the estimators is here the 
%quantity
object 
of interest. 
%So, in this scenario, the corresponding relations are given in terms of matrix inequalities instead of scalar inequations. 
%%%%%% apototo bad%%%%%%%%%%%%%%%%%
%The simultaneous estimation of independent local fields of a many-body encoding Hamiltonian is an arena which is hitherto unexplored.
%%%%%%%%%%%%%%%%%%%%%%%%%%%%%%%
We find that even though the parameters to be estimated are independent local ones,
genuine multipartite entanglement is indispensable in the optimal input probe.
%%%%%%%%%% apototo bad %%%%%%%%%%%%%%%%%%%%%%%%
% \textit{only} the Greenberger-Horne-Zeilinger (GHZ)~\cite{ghz1,ghz2}
% %\textcolor{magenta}{[GHZ conf proc.;  Mermin Am J Phys]} 
% state, up to a relative phase, in the eigenbasis of the encoding Hamiltonian,  is the optimal probe.
%%%%%%%%% apototo bad %%%%%%%%%%%%
% To measure the precision of estimation of multiple parameters, and to compare with other instances, it is usually convenient to consider a trace of the matrix inequalities after taking a product with a positive semi-definite weight matrix. The minimal value of this quantity provides the best metrological precision in multiparameter estimation for the chosen weight matrix~\cite{rev0,ref9,dur}.
%%%%%%%%%%%%%%%%%%%%%%%%%%%%%%%%%%%%%%%%%%%%%%%%%%
%Along the way of proving our main results, we provide an axiomatic formulation of why one can choose this quantity as a ``figure of merit" in quantum multiparameter estimation. 
%%%%
%%%%We prove the result about optimality and uniqueness of the GHZ probe in three steps. 
Our results comprise three main steps.
First, we optimize the quantum Cram\'er-Rao bound with respect to pure probes, and provide a general lower bound in terms of an arbitrary positive semi-definite weight matrix.
%of multiparameter estimation.  
The bound holds for an arbitrary local encoding Hamiltonian and for an arbitrary number of parties, each having arbitrary dimension.
%Then we show that for a particular choice of the weight matrix and for a given local encoding Hamiltonian, the bound is 
%a constant, 
%independent of the number of parties. 
%In the second step, we prove that, when optimized over pure input probes, the GHZ state, up to a relative phase, is the only state that attains the bound.
In the second step, we prove that, genuine multipartite entanglement is a prerequisite to achieve the optimal precision, and there always exits a GHZ state in a particular basis that saturates the bound.
Finally, in the third step, we have  scanned the space of all states, including mixed ones,  and shown that no mixed state provides a precision better than that provided by any pure state.
We also provide the parametric form of the most general optimal state for three parties.
%the GHZ state.
%up to a relative phase. 
Furthermore, we find the best precision attainable with pure product probes.

Experiments have been performed which are concerned with the construction of quantum computers utilizing one- or two-dimensional arrays of spins, particularly using ion-traps or micro-traps~\cite{q1,q2,q3,q4,q5,q6,q7}. Such arrays of spins, when actually present in the laboratory, cannot be completely isolated as other devices present in the laboratory inevitably affect these spin-arrays. As a result, the arrays of spins, necessary for constructing quantum computers, are affected by stray fields. Such stray fields affect different spins differently depending upon the relative positions of the different spins in the lattice with respect to the external devices in the laboratory. For instance, the spins near the boundary are affected differently than the spins present in the bulk of the lattice. Estimating such stray fields is essential in order to combat the effects of nearby apparatus in the laboratory and for efficient functioning of the quantum computer. Since these stray fields affect different spins differently, estimating such field strengths essentially reduces the problem to an estimation of multiple independent field strengths of the relevant Hamiltonian. This explains how the multiparameter estimation of independent field strengths can potentially find its application in realistic experiments.\\
\section{Multiparameter estimation}
\label{adbhut_andhare}
%%%%%%%%%%%%%%%%%-ref-%%%%%%%%%%%%%%%%%%%
%\textcolor{blue}{A discussion on multiparameter estimation is provided in the Supplementary material (SM).}
%%%%%%%%%%%%%%%%%%%%%%%%%%%%%%%%%%%%%
% \textcolor{magenta}{The quantum Fisher information (QFI)-based figure of merit, generally considered for multiparameter estimation, is given by $\Tr(WF_Q^{-1})$, where $F_Q$ is the QFIM, and $W$ is an arbitrary positive semi-definite operator~\cite{ref11}.
% %%%%%%%%%%%%%%%%%%%-ref-%%%%%%%%%%%%%%%%%%%%
% %\textcolor{blue}{Here we provide the axioms that are satisfied by the quantity, $\Tr(WF_Q^{-1})$, which certifies its use as a valid figure of merit for multiparameter estimation, where $F_Q$ is the quantum Fisher information matrix (QFIM) and $W$ is an arbitrary positive semi-definite weight matrix. The two axioms are as follows.} 
% %%%%%%%%%%%%%%%%%%%%%%%%%%%%%%%%%%%%
% The 
% %QFI-based figure of merit, 
% quantity, $\Tr(WF_Q^{-1})$, is positive, 
% %, i.e. $\Tr(WF_Q^{-1})\ge 0$, 
% and is monotonic under the action of completely positive trace preserving (CPTP) operations.
% %, i.e. 
% %%%%%%%%%%%%% apototo baad %%%%%%%%%%%%%%%%%%%
% %This means that $\Tr\left[WF_Q(\rho_{\phi})^{-1}\right] \le \Tr\left[WF_Q(\Lambda({\rho_{\phi}}))^{-1}\right]$, where $\Lambda$ is an arbitrary CPTP map acting on the encoded probe, $\rho_{\phi}$. 
% %%%%%%%%%%%%%%%%%%%%%%%%%%%%%%%%%%%%%%%%%%%%%
% The proofs of 
% %of the axioms 
% these properties and a detailed discussion on multiparameter estimation are provided in the SM.}
The aim of multiparameter estimation is to estimate a set of parameters, encoded onto an input probe, by performing suitable measurements $M_x$ on the encoded probe.
%Let a set of parameters, $\phi=\{\phi_1,\phi_2,...,\phi_n\}$, be encoded onto an input probe, $\rho$, using some physical process. The aim is to estimate these parameters with best achievable precision by performing suitable measurements, $M_x$, on the encoded probe, $\rho_{\phi}$. 
The probabilities of the measurement outcomes, $x$, belong to to the distribution, $p(x|\phi)=\Tr(\rho_{\phi}M_x)$, where $\rho_{\phi}$ denotes the encoded state. The problem of estimation of multiple parameters is formulated in terms of a covariance matrix, defined by $\Cov(\hat{\phi})_{ij}=\langle(\hat{\phi}_i-\phi_i)(\hat{\phi}_j-\phi_j) \rangle$, $\forall i,j$ where the expectation is taken with respect to the probability distribution, $p(x|\phi)$.
One of the fundamental bounds in multiparameter estimation is given by $\nu \Tr(W \Cov(\hat{\phi}))\ge \Tr(WF^{-1})$, for an arbitrary positive semi-definite matrix, $W$, and $\nu$ repetitions of the experiment.
Here $F$ is the Fisher information matrix, which depends on the measurement setting and the choice of the input probe.
The right hand side of this inequality
%, $\nu \Tr(W \Cov(\hat{\phi}))\ge \Tr(WF^{-1})$, %\textcolor{magenta}{depends on the choice of probe state and the measurements performed. In principle, one }
can be minimized over all measurements to obtain the quantum Cram\'er-Rao bound, which is given in terms of the quantum Fisher information matrix (QFIM), denoted by $F_Q$. The corresponding quantum Cram\'er-Rao inequality is given by $ \Tr(WF^{-1}) \ge \Tr(WF_Q^{-1})$, where $W$ is an arbitrary positive semi-definite weight matrix.
% \textcolor{magenta}{\begin{eqnarray}
% \label{alo_chhaya}
%     F^{-1} \ge F_Q^{-1}.
% \end{eqnarray}}
%\textcolor{orange}{Inequation~\eqref{alo_chhaya} is again a matrix inequality, which means that the matrix, $F^{-1} - F_Q^{-1}$, is positive semi-definite.}\textcolor{cyan}{lekhar dorkar achhe?} 
The elements of QFIM, $F_Q$, are defined as $(F_Q)_{ij}=\Tr(\rho_{\phi}\{L_i,L_j\})/2, \; \forall i,j$, where $L_i$ is a Hermitian matrix,  referred to as the symmetric logarithmic derivative (SLD) corresponding to the encoded state, $\rho_{\phi}$.
If the encoded state is pure, say $\ket{\psi_{\phi}}$, then the corresponding SLD operator for the $i^{\text{th}}$ parameter is 
$L_i=2(\ketbra{\partial_{\phi_i}\psi_{\phi}}{\psi_{\phi}}+\ketbra{\psi_{\phi}}{\partial_{\phi_i}\psi_{\phi}})$.
%%%%%%%%%%%%% apototo baad %%%%%%%%%%%%%%%%%%%%%%%%%%
% \textcolor{magenta}{\begin{eqnarray}
% (F_Q)_{ij}=\frac{1}{2}\Tr(\rho_{\phi}\{L_i,L_j\}), \;\;\; \forall i,j
% \end{eqnarray}
% where the curly brackets denote anti-commutator, and $L_i$ is a Hermitian matrix,  referred to as the symmetric logarithmic derivative (SLD) corresponding to the encoded state, $\rho_{\phi}$. It is defined using the relation $\frac{1}{2}(L_i \rho_{\phi}+\rho_{\phi}L_i)=\partial_{\phi_i}\rho_{\phi}$, for any encoded $\rho_{\phi}$. It is to be noted that the QFIM is real and symmetric, i.e. $(F_Q)_{ij}=(F_Q)_{ji}$. The SLD operator corresponding to the $i^{\text{th}}$ parameter is explicitly given by
% \begin{eqnarray}
% \label{jharna_dhara}
% L_i=2\sum_{a,b}\frac{\bra{\psi_a}\partial_{\phi_i}\rho_{\phi}\ket{\psi_b}}{\lambda_a+\lambda_b} \ketbra{\psi_a}{\psi_b},
% \end{eqnarray}
% where the encoded state is written in spectral decomposition as $\rho_{\phi}=\sum_k \lambda_k \ketbra{\psi_k}{\psi_k}$. If the encoded state is pure, say $\ket{\psi_{\phi}}$, then the corresponding SLD operator for the $i^{\text{th}}$ parameter further simplifies to 
% \begin{eqnarray}
% \label{brisTi}
% L_i=2(\ketbra{\partial_{\phi_i}\psi_{\phi}}{\psi_{\phi}}+\ketbra{\psi_{\phi}}{\partial_{\phi_i}\psi_{\phi}}).
%\end{eqnarray}}
%%%%%%%%%%%%%%%%%%%%%%%%%%%%%%%%%%%%%%%%%%%%%%
%The diagonal elements of the QFIM are given by $(F_Q)_{ii}=\Tr(\rho_{\phi}L_i^2)$, which essentially provides the quantum Fisher information for the $i^{\text{th}}$ parameter. 
The quantum Fisher information (QFI)-based figure of merit, generally considered for multiparameter estimation, is given by $\Tr(WF_Q^{-1})$, where $F_Q$ is the QFIM, and $W$ is an arbitrary positive semi-definite operator. This figure of merit is positive, and monotonic under completely positive trace preserving operations~\cite{rev2014,rev2015}. \\

\section{Estimation of independent local fields  using pure probes}
\label{mohini}
We consider unitary encoding of input probes. 
The encoding unitary operator is given by $U=\exp(-i\kappa \widetilde{H}t/\hbar)$, where $\kappa$ has magnitude one and the dimension of $time^{-1}$. We observe the system at time, $t=1$. Therefore the unitary operator effectively is given by $U=\exp(-i\widetilde{H}/\hbar)$. The generator of the unitary contains an arbitrary $N$-party encoding Hamiltonian, $\widetilde{H}$, containing single-party terms. The Hamiltonian, $\widetilde{H}$ is given by $\widetilde{H}=\hbar \sum_{i=1}^N h_i H_i$, where $H_i$ are the local Hamiltonians corresponding to the $i^{\text{th}}$ party, having arbitrary dimension.
Our aim is to simultaneously estimate all the independent field strengths, $h_i$, corresponding to every local term in the Hamiltonian. 
In theorem~\ref{bound_gen}, we minimize the QFI-based figure of merit, given by $\Tr(W F_Q^{-1})$, with respect to all $N$-party pure input probes  for an arbitrary choice of positive semi-definite $W$, and an arbitrary encoding many-body Hamiltonian, $\widetilde{H}$, comprising of a sum of single-body terms. 
%\textcolor{red}{Thus, although the figure of merit depends on the choice of the input probe, the bound that we obtain in theorem~\ref{bound_gen} depends only on the arbitrary positive semi-definite weight matrix, $W$, and the encoding Hamiltonian $\widetilde{H}$, having arbitrary local components, $H_i$.}
We then find the actual value of the bound obtained in theorem~\ref{bound_gen}, while pertaining to a particular Hamiltonian $H=\hbar \sum_{i=1}^N  h_i \sigma_z^i$ , and a full-rank positive semi-definite weight matrix,  $\overline{W}=\sum_{i=1}^{N}a\ketbra{i}{i}+\sum_{i,j=1, i\ne j}^{N}b\ketbra{i}{j}$, where 
$a=16[(N-1)\alpha^2+1]$, and $b=16[(N-2)\alpha^2+2\alpha]$, for $0<\alpha<1$ and a given value of $N$. The weight matrix $\overline{W}$ is full-rank, with non-zero off-diagonal elements. 
%%%%%%%%%%%%%%%%%%%%%%%%%%%%%%%%%%%%%%%%%%%%%%%%%
Every diagonal element of $\overline{W}$ is greater than each off-diagonal element, i.e. $a>b$, which assures the positive semi-definiteness of the matrix.
%%%%%%%%%%%%%%%%%%%%%%%%%%%%%%%%%%%%%%%%%%%%%
%$a,b>0$ and $a>b$. 
Further in Corollary~\ref{pahar}, we provide the necessary and sufficient condition of attainability of the bound in theorem~\ref{bound_gen} generally for arbitrary $W$ and Hamiltonian, $\widetilde{H}$, and then exemplify the case with weight matrix $\overline{W}$ and Hamiltonian $H$.
%\textcolor{magenta}{Then we pertain to a special choice of the Hamiltonian, $H=\hbar \sum_{i=1}^N  h_i \sigma_z^i$, and weight matrix, $\mathcal{W}=\ketbra{\mathcal{W}}{\mathcal{W}}$, where $\ket{\mathcal{W}}=\sum_{i=0}^{N-1} \ket{i}$ in the eigenbasis of $H$. We show that specifically  for  $W=\mathcal{W}$ and $F_Q=\mathcal{F}_Q$, the lower bound is a constant and  independent of the value of the number of parties. Here $\mathcal{F}_Q$ is the QFIM corresponding to the encoding Hamiltonian, $H$, and pure encoding probes.Additionally, we find the condition of saturability of the lower bound that we provide. Further, still considering the weight matrix to be $\mathcal{W}$ and the QFIM to be $\mathcal{F}_Q$, we find that optimal precision is attainable using only the Greenberger-Horne-Zeilinger state, up to a relative phase, when the optimization is performed over all pure states. We next find the minimum value of the QFI-based figure of merit when the optimization is performed over all pure product states. }
It is to note that the quantum Cram\'er-Rao bound is attainable in the scenario that we consider, since in this case, $\bra{\psi_0}[L_i,L_j]\ket{\psi_0}=0$, which happens to be a necessary and sufficient condition of saturability of the multiparameter quantum Cram\'er-Rao bound~\cite{Matsumoto,Rafal}.
The elements of the QFIM, $\mathcal{F}_Q$, corresponding to the encoding Hamiltonian, $H$, are given by
\begin{eqnarray}
\label{dakhshini}
 (\mathcal{F}_Q)_{ii} &=& 4\left[1-\bra{\psi_0}\sigma_z^i \ket{\psi_0}^2 \right], \;\; \forall i \\
 \label{vrantibilas}
 (\mathcal{F}_Q)_{ij} &=& 4\Big[\bra{\psi_0} \sigma_z^i  \otimes \sigma_z^j  \ket{\psi_0} \nonumber \\
 &-&  \bra{\psi_0} \sigma_z^i  \ket{\psi_0}\bra{\psi_0} \sigma_z^j \ket{\psi_0}\Big] \;\; \forall i,j, i\ne j 
 % &=& (\mathcal{F}_Q)_{ji}, \nonumber
\end{eqnarray}
where each of $i$ and $j$ can take values $1$ to $N$, and $\ket{\psi_0}$ denotes the input probe.

\begin{theorem}
\label{bound_gen}
The lower bound of the QFI-based figure of merit, $\Tr(WF_Q^{-1})$, corresponding to the $N$-party encoding Hamiltonian, $\widetilde{H}=\hbar \sum_{i=1}^N h_i H_i$, which contains only single-body terms, $H_i$, is given by
%In the estimation of $N$ independent single-party field strengths, given by  $h=\{h_1,h_2,...,h_N\}$, of an $N$-party encoding Hamiltonian, $\widetilde{H}=\hbar \sum_{i=1}^N h_i H_i$, containing only one-body terms, and considering an arbitrary weight matrix, $W$, the QFI-based figure of merit, $\Tr(W\widetilde{F}_Q^{-1})$, has a lower bound given by
\begin{eqnarray}
\label{banalata}
\Tr(WF_Q^{-1}) \ge \frac{\Tr(W^{1/2})^2}{\sum_{i=1}^N(\lambda_M^i-\lambda_m^i)^2},
\end{eqnarray}
where $W$ is an arbitrary positive semi-definite weight matrix,  $\lambda_{M(m)}^i$ denotes the maximum (minimum) eigenvalue of each local term $H_i$, and the minimization is performed over all $N$-party pure input probes. 
\end{theorem}

\noindent \textit{Proof.} %Let us consider that the inverse of the QFIM, $F_Q$, exists, and let $W$ be an arbitrary positive semi-definite weight matrix. Then we obtain the following chain of inequalities:
For an invertible QFIM $F_Q$, and an arbitrary positive semi-definite weight matrix, $W$,  we obtain the following chain of inequalities:
\begin{eqnarray}
\label{falguni}
    && \Tr(W^{1/2}) = \Tr\left(F_Q^{1/2}W^{1/2} F_Q^{-1/2}\right) \nonumber \\
    &=& \Tr\left[\left(F_Q^{1/2}\right)^{\dagger}W^{1/2} F_Q^{-1/2}\right] \nonumber \\
    &\le & \sqrt{\Tr\left[\left(F_Q^{1/2}\right)^{\dagger}F_Q^{1/2}\right]} \nonumber \\
   && \sqrt{\Tr\left[\left(W^{1/2} F_Q^{-1/2}\right)^{\dagger} W^{1/2} F_Q^{-1/2}\right]}\;\;\; \nonumber \\
   %&=& \sqrt{\Tr(F_Q) \Tr\left[W^{1/2} F_Q^{-1}W^{1/2}\right]} \nonumber \\
    & = & \sqrt{\Tr(F_Q) \Tr(F_Q^{-1}W)}. 
\end{eqnarray}
In deriving the above inequalities, 
%%%%%%%%%%%%%%%%%%%%%%%%%%%%%%%%%%%%%%%%%%%%%%%%%%%%%%%%
%%%%%%%%%%%%%%%%%% apototo baad %%%%%%%%%%%%%%%%%%%%%%%%%%%%%%%%%%%%%%
% we have used the property
% there exists one positive semi-definite hermitian matrix which is the square root of a positive semi-definite hermitian matrix.
% Specifically, in the second step we have used the feature that $F_Q^{1/2}$ is Hermitian, since $F_Q$
% positive semi-definite.
%In the third step, 
%%%%%%%%%%%%%%%%%%%%%%%%%%%%%%%%%%%%%%%%%%%%%%%%%%%%%%%%
%%%%%%%%%%%%%%%%%%%%%%%%%%%%%%%%%%%%%%%%%%%%%%%%%%%%%%%%
we have used the Cauchy-Schwarz inequality in the third step.
%\textcolor{blue}{\sout{, and the fact that $\Tr(W^{1/2})$ is a real quantity since $W^{1/2}\ge 0$}}
The Cauchy-Schwarz inequality for any two operators, $A$ and $B$, is given by $|\Tr(A^{\dagger}B)|^2\le \Tr(A^{\dagger}A)\Tr(B^{\dagger}B)$. The fourth step follows since both $F_Q^{-1/2}$ and  $W^{1/2}$ are hermitian. % and positive semi-definite 
%\textcolor{blue}{\sout{, and hence are Hermitian}}
Therefore from inequation~\eqref{falguni}, we obtain the inequality
\begin{eqnarray}
\label{gantabya}
    \Tr(WF_Q^{-1})\ge \frac{\Tr(W^{1/2})^2}{\Tr(F_Q)}.
\end{eqnarray}
Now, our aim is to minimize the quantity $\Tr(W^{1/2})^2/\Tr(F_Q)$ 
%the quantity on the right hand side of the above inequality, 
over all pure input probes, for a given arbitrary positive semi-definite weight matrix $W$. This essentially reduces the problem to a  maximization of the function  $\Tr(F_Q)$ over pure input states, for an encoding Hamiltonian, $\widetilde{H}=\hbar \sum_{i=1}^N h_i H_i$.
 %Consider the encoding Hamiltonian to be $\widetilde{H}=\hbar \sum_{i=1}^N h_i H_i$, where 
 %Our aim is to simultaneously estimate $h_i$, for $i=1$ to $N$. 
 Let the pure input probe that maximizes $\Tr(F_Q)$ be denoted by $\ket{\widetilde{{\psi}_0}}$. If the corresponding QFIM be denoted by $\widetilde{F}_Q$, then $ \Tr(\widetilde{F}_Q) = 4\sum_i^N \left[\bra{\widetilde{{\psi}_0}}H_i^2 \ket{\widetilde{{\psi}_0}}- \bra{\widetilde{{\psi}_0}}H_i \ket{\widetilde{{\psi}_0}}^2 \right]$, which can be written
% \begin{eqnarray}
% \label{kakhono_megh}
%    \Tr(\widetilde{F}_Q) = 4\sum_i^N \left[\bra{\widetilde{{\psi}_0}}H_i^2 \ket{\widetilde{{\psi}_0}}- \bra{\widetilde{{\psi}_0}}H_i \ket{\widetilde{{\psi}_0}}^2 \right].
% \end{eqnarray}
%One can further write this equation 
%Eq.~\eqref{kakhono_megh} 
as
\begin{eqnarray}
\label{lal_kamal}
     \Tr(\widetilde{F}_Q) &=& 4\sum_{i=1}^N  \left[\Tr(H_i^2\widetilde{\rho_i})-\Tr(H_i \widetilde{\rho_i})^2\right] 
     = 4\sum_{i=1}^N \Delta^2_{\rho_i} H_i, \nonumber
\end{eqnarray}
where $\widetilde{\rho_i}=\Tr_{
\overline{i}}(\ketbra{\widetilde{{\psi}_0}}{\widetilde{{\psi}_0}})$, and $\overline{i}$ denotes every other party except the $i^{\text{th}}$ one.  Here $\widetilde{\rho_i}$ is the reduced density matrix corresponding to the $i^{\text{th}}$ party, which can be, in general, mixed. 
%Since $\ket{\widetilde{\psi_0}}$ is the state that maximizes $\Tr(F_Q)$, $\Tr(\widetilde{F}_Q)$ corresponds to the maximum of $\Tr(F_Q)$.
Now, if we consider each term of the summation in $\Tr(\widetilde{F}_Q)$ in the above equation, %Eq.~\eqref{lal_kamal},
%, i.e. $\Delta^2_{\rho_i} H_i$, 
then the maximum of every such term is attained by pure states, where the maximum value is given by  $(\lambda_M^i-\lambda_m^i)^2/4$~\cite{variance}. The corresponding optimal state that maximizes $\Delta^2_{\rho_i} H_i$ is $\ket{\chi^i}=(\ket{\lambda_M^i}+\ket{\lambda_m^i})/\sqrt{2}$, $\forall i$, where $\ket{\lambda_{M(m)}^i}$ denotes the eigenvectors corresponding to the maximum (minimum) eigenvalue of $H_i$. Therefore the state  that maximizes $\Tr(\widetilde{{F}_Q})$ is $\ket{\widetilde{\psi_0}}=\ket{\chi^i}^{\otimes N}$, which is pure product.
From the above analysis, we find that the maximum value of $\Tr(\widetilde{F}_Q)$ is $\sum_{i=1}^N(\lambda_M^i-\lambda_m^i)^2$, when optimized over pure probes. Using inequation~\eqref{gantabya}, this finally leads us to the result that the minimum value of $\Tr(WF_Q^{-1})$, when optimized over all pure input probes,  is given by $\Tr(W^{1/2})^2/\xi$, where $\xi=\sum_{i=1}^N(\lambda_M^i-\lambda_m^i)^2$, and $W$ is an arbitrary positive semi-definite matrix, which proves theorem~\ref{bound_gen}. \hfill $\blacksquare$
\\

If the Hamiltonian considered is $H=\hbar \sum_{i=1}^N  h_i \sigma_z^i$, and the weight matrix is $\overline{W}$, the value of the lower bound obtained in theorem~\ref{bound_gen} is given by $4N$. A remark on Theorem~\ref{bound_gen} is provided in~\ref{appen1}.

\begin{corollary}
\label{pahar}
The necessary and sufficient condition of attainability of the inequation $\Tr(WF_Q^{-1}) \ge \Tr(W^{1/2})^2/\Tr(F_Q)$ is given by 
\begin{eqnarray}
\label{oporajito}
    F_Q^{1/2}=\frac{\Tr({F_Q})}{\Tr({W^{1/2}})}F_Q^{-1/2}W^{1/2},
\end{eqnarray}
for an arbitrary positive semi-definite matrix, $W$, and QFIM $F_Q$. Specifically for full-rank positive semi-definite weight matrix $\overline{W}$, and the Hamiltonian, $H=\hbar \sum_{i=1}^N  h_i \sigma_z^i$, the equality condition becomes $F_Q=4\left(\sum_{i=1}^{N}\ketbra{i}{i}+\sum_{i,j=1, i\ne j}^{N}\alpha \ketbra{i}{j}\right)$, where $0 < \alpha <1$.  Here $\overline{W}=\sum_{i=1}^{N}a\ketbra{i}{i}+\sum_{i,j=1, i\ne j}^{N}b\ketbra{i}{j}$, where 
$a=16[(N-1)\alpha^2+1]$, and $b=16[(N-2)\alpha^2+2\alpha]$, for a given value of $N$.
\end{corollary}
%
% \begin{corollary}
% \label{pahar}
% The minimum value of $\Tr(\mathcal{W}(\mathcal{F}_Q)^{-1})$ for an encoding Hamiltonian, $H=\hbar \sum_{i=1}^N  h_i \sigma_z^i$, is attained by the pure probe state which satisfies the condition, $\lim_{(\alpha,\beta)\to (1,1)} \mathcal{F}_Q = 4\mathcal{W}$, where $\mathcal{F}_Q$ is 
% the corresponding QFIM, 
% %$\mathcal{\widetilde{F}}_Q$ is a positive semi-definite matrix 
% whose each diagonal and off-diagonal element is $4\alpha$ and $4\beta$ respectively, 
% and $\mathcal{W}=\ketbra{\mathcal{W}}{\mathcal{W}}$, where $\ket{\mathcal{W}}=\sum_{i=0}^{N-1} \ket{i}$ in the computational basis.
% \end{corollary}

\noindent \textit{Proof.} The equality in $\Tr(WF_Q^{-1}) \ge \Tr(W^{1/2})^2/\Tr(F_Q)$ follows from the equality condition of the Cauchy-Schwarz inequality used to derive inequation~\eqref{falguni}. It is a necessary and sufficient condition of equality. The equality condition of $|\Tr(A^{\dagger}B)|^2\le \Tr(A^{\dagger}A)\Tr(B^{\dagger}B)$ is given by $A=\widetilde{K}B$, where $\widetilde{K}$ is a real or complex number. In the scenario that we consider, the corresponding condition becomes $F_Q^{1/2}=\widetilde{K}F_Q^{-1/2}W^{1/2}$. This is equivalent to $F_Q=\widetilde{K}W^{1/2}$, where $\widetilde{K}$ is a real number equal to $\Tr(F_Q)/\Tr(W^{1/2})$. Thus we obtain Eq.~\eqref{oporajito}. Next we find the value of $\widetilde{K}$ for the particular weight matrix $\overline{W}$ and Hamiltonian $H$. Considering $W=\overline{W}$ and the Hamiltonian, $H$, we already found that the value of the lower bound in theorem~\ref{bound_gen} is $4N$. Now, if we put the equality condition in the left hand side of the bound, i.e. in $\Tr(WF_Q^{-1})$, we obtain $\Tr(\overline{W}\overline{W}^{-1/2}\widetilde{K}^{-1})=\widetilde{K}^{-1}\Tr(\overline{W}^{1/2})=\widetilde{K}^{-1}4N$. The last equality follows from the fact that $\overline{W}^{1/2}=4\left(\sum_{i=1}^{N}\ketbra{i}{i}+\sum_{i,j=1, i\ne j}^{N}\alpha \ketbra{i}{j}\right)$, for all $0 < \alpha <1$. For details of the calculation of $\overline{W}^{1/2}$, refer to Appendix~\ref{appen}. Therefore we obtain $4N=\widetilde{K}^{-1}4N$, which gives $\widetilde{K}=1$. Hence for $W=\overline{W}$ and the Hamiltonian, $H$, the equality condition is $F_Q=\overline{W}^{1/2}=4\left(\sum_{i=1}^{N}\ketbra{i}{i}+\sum_{i,j=1, i\ne j}^{N}\alpha \ketbra{i}{j}\right)$, which proves Corollary~\ref{pahar}.
\hfill $\blacksquare$ \\

We have seen previously that the quantum Cram\'er-Rao bound for the Hamiltonian $H$ is achievable for all choices of input probes. Further, 
we optimize the quantum Cram\'er-Rao bound over input probes, and find that 
the lower bound thus obtained 
%in theorem~\ref{bound_gen} 
is also attainable
%, as observed in 
(Corollary~\ref{pahar}). Specifically, for the weight matrix, $\overline{W}$, this lower bound is achievable at $F_Q=\overline{W}^{1/2}$. In the succeeding theorem~\ref{gme_new}, we find that genuine multipartite entanglement is necessary in the input probe to attain the bound $F_Q=\overline{W}^{1/2}$. Next we show in theorem~\ref{ghz_new} that there always exists at least one  GHZ state which saturates this bound.

\begin{theorem}
\label{gme_new}
The optimal pure input probe, which attains  the lower bound of the quantity $\Tr(WF_Q^{-1})$, for the Hamiltonain, $H=\hbar \sum_{i=1}^N  h_i \sigma_z^i$, and full-rank positive semi-definite weight matrix, $W=\overline{W}$, is necessarily genuine multipartite entangled. Here $\overline{W}=\sum_{i=1}^{N}a\ketbra{i}{i}+\sum_{i,j=1, i\ne j}^{N}b\ketbra{i}{j}$, where 
$a=16[(N-1)\alpha^2+1]$, and $b=16[(N-2)\alpha^2+2\alpha]$, $\forall \alpha$ in the range $0<\alpha<1$ and a given value of $N$.
\end{theorem}

\noindent \textit{Proof.} The value of the lower bound in theorem~\ref{bound_gen} is given by $4N$, for the Hamiltonian $H$ and weight matrix $\overline{W}$. We found in corollary~\ref{pahar} that this is achieved when the condition, 
\begin{eqnarray}
    F_Q=4\left(\sum_{i=1}^{N}\ketbra{i}{i}+\sum_{i,j=1, i\ne j}^{N}\alpha\ketbra{i}{j}\right),
\end{eqnarray}
is satisfied.
Using this condition in equations~\eqref{dakhshini} and~\eqref{vrantibilas}, we obtain $\bra{\psi_0}\sigma_z^i \ket{\psi_0}=0$, and $\bra{\psi_0} \sigma_z^i  \otimes \sigma_z^j  \ket{\psi_0}=\alpha$, $\forall$ $0<\alpha<1$. Now let us suppose that the optimal pure input probe is product in at least one bipartition. So the input state is of the form $\ket{\psi_0}=\ket{\psi_i}\otimes\ket{\psi_j}$, where the notation $\ket{\psi_i}$ denotes that the site  at one end of the bipartition is the $i^{\text{th}}$ one. This implies $\bra{\psi_0} \sigma_z^i  \otimes \sigma_z^j  \ket{\psi_0}= \bra{\psi_i \otimes \psi_j} \sigma_z^i  \otimes \sigma_z^j  \ket{\psi_i \otimes \psi_j}=\bra{\psi_i}\sigma_z^i \ket{\psi_i}\bra{\psi_j}\sigma_z^i \ket{\psi_j}=0$, which is clearly a contradiction. This argument follows for all values of $i$ and $j$. Hence our assumption that the pure input probe is product in at least one bipartition is wrong. This proves that the optimal input probe is necessarily genuine multipartite entangled. \hfill $\blacksquare$ 
\begin{theorem}
\label{ghz_new}
There always exists a Greenberger-Horne-Zeilinger state, with a fixed relative phase, which saturates the inequality $\Tr(WF_Q^{-1}) \ge \Tr(W^{1/2})^2/\Tr(F_Q)$ for the Hamiltonian, $H=\hbar \sum_{i=1}^N  h_i \sigma_z^i$, and for every choice of the full-rank positive semi-definite weight matrix, $W=\overline{W}$. Here $\overline{W}=\sum_{i=1}^{N}a\ketbra{i}{i}+\sum_{i,j=1, i\ne j}^{N}b\ketbra{i}{j}$, where 
$a=16[(N-1)\alpha^2+1]$, and $b=16[(N-2)\alpha^2+2\alpha]$, $\forall \alpha$ belonging to $0<\alpha<1$ and a given value of $N$. The explicit form of the GHZ state in the computational basis is $\left(\ket{\psi}^{\otimes N}+e^{i\phi}\ket{\psi^{\perp}}^{\otimes N}\right)/\sqrt{2}$, where $\ket{\psi}=\cos{(\theta/2)}\ket{0}+\sin{(\theta/2)}\ket{1}$ and $\ket{\psi^{\perp}}=-\sin{(\theta/2)}\ket{0}+\cos{(\theta/2)}\ket{1}$, with the values of the parameters, $\theta=\cos^{-1}(\pm \sqrt{\alpha})$ and $\phi=(2n+1)\pi/2$, for $n$ being a non-negative integer. 
\end{theorem}

\noindent \textit{Proof.} 
In Corollary~\ref{pahar}, we found that the equality in $\Tr(WF_Q^{-1}) \ge \Tr(W^{1/2})^2/\Tr(F_Q)$ is satisfied when $F_Q=4\left(\sum_{i=1}^{N}\ketbra{i}{i}+\sum_{i,j=1, i\ne j}^{N}\alpha\ketbra{i}{j}\right)$, for the Hamiltonian $H$ and weight matrix $\overline{W}$. 
Using this condition in equations~\eqref{dakhshini} and~\eqref{vrantibilas}, we obtain $\bra{\psi_0}\sigma_z^i \ket{\psi_0}=0$, and $\bra{\psi_0} \sigma_z^i  \otimes \sigma_z^j  \ket{\psi_0}=\alpha$, $\forall$ $0<\alpha<1$.
Let us now consider the input probe to be of the form $\ket{\overline{\psi}}=\left(\ket{\psi}^{\otimes N}+e^{i\phi}\ket{\psi^{\perp}}^{\otimes N}\right)/\sqrt{2}$, where $\ket{\psi}=\cos{(\theta/2)}\ket{0}+\sin{(\theta/2)}\ket{1}$ and $\ket{\psi^{\perp}}=-\sin{(\theta/2)}\ket{0}+\cos{(\theta/2)}\ket{1}$. Our aim is to find the values of $\theta$ and $\phi$ that satisfy equations~\eqref{dakhshini} and~\eqref{vrantibilas}. Eq.~\eqref{dakhshini} can be written as $\bra{\overline{\psi}}\sigma_z\ket{\overline{\psi}}=-\sin\theta \cos \phi=0$, while Eq.~\eqref{vrantibilas} becomes $\bra{\overline{\psi}}\sigma_z^i\otimes\sigma_z^j\ket{\overline{\psi}}=\cos^2\theta+\sin^2\theta\cos\phi=\alpha$. The solution of these two equations is given by $\theta=\cos^{-1}(\pm \sqrt{\alpha})$ and $\phi=(2n+1)\pi/2$, where $n$ is a non-negative integer. 
%There exits a unique solution to the two equations,  $\bra{\overline{\psi}}\sigma_z\ket{\overline{\psi}}=0$ and $\bra{\overline{\psi}}\sigma_z^i\otimes\sigma_z^j\ket{\overline{\psi}}=\alpha$, for every value of $\alpha$ where $0<\alpha<1$. 
Therefore the GHZ state corresponding to $\theta=\cos^{-1}(\pm \sqrt{\alpha})$ and $\phi=(2n+1)\pi/2$, with $n$ being a non-negative integer, corresponds to the optimal probe which saturates the bound in theorem~\ref{bound_gen} for the Hamiltonian, $H$ and weight matrix, $\overline{W}$.  \hfill $\blacksquare$ \\

The parametric form of the most general pure input state, for the scenarios when $N$ is odd, denoted by $N_o$, and when $N$ is even, denoted by $N_e$, are respectively given by
\begin{eqnarray}
\ket{\psi_o}&=&\alpha_0 \ket{0}^{\otimes N_o} + \overline{\alpha_0} \ket{1}^{\otimes N_o}  + \sum_{i=1}^{(N_o-1)/2}\sum_{j=1}^{\binom{N_o}{i}} \alpha_{ij}\ket{W_{ij}} \nonumber \\
    &+& \sum_{i=1}^{(N_o-1)/2}\sum_{j=1}^{\binom{N_o}{i}} \overline{\alpha_{ij}}\ket{\overline{W}_{ij}}  \;\;\; \text{and} \nonumber \\
    \ket{\psi_e}&=&\alpha_0 \ket{0}^{\otimes N_e} + \overline{\alpha_0} \ket{1}^{\otimes N_e}  + \sum_{i=1}^{N_e/2-1}\sum_{j=1}^{\binom{N_e}{i}} \alpha_{ij}\ket{W_{ij}} \nonumber \\
    &+& \sum_{i=1}^{N_e/2-1}\sum_{j=1}^{\binom{N_e}{i}} \overline{\alpha_{ij}}\ket{\overline{W}_{ij}} + \sum_{j=1}^{\binom{N_e}{(N_e/2) j}}\alpha_{\frac{N_e}{2}j}\ket{W_{\frac{N_e}{2}} j},\nonumber
\end{eqnarray}
where $\ket{W_{ij}} (\ket{\overline{W}_{ij}})$ denotes the $W$-state with $i$ down (up) spins $\left(\ket{1(0)}\right)$, and $j$ is the number of such $W$-states corresponding to each value of $i$. The state, $\ket{\psi_{o(e)}}$, in the respective cases, satisfying the two conditions, $\bra{\psi_{o(e)}}\sigma_z\ket{\psi_{o(e)}}=0$ and $\bra{\psi_{o(e)}}\sigma_z^i\otimes\sigma_z^j\ket{\psi_{o(e)}}=\alpha$, is the most general input probe. In particular, for $N=3$, the parametric form of the input state corresponds to
\begin{eqnarray}
    |\overline{\alpha_0}|^2 &=& \frac{1+3\alpha}{4}-\sin^2\chi, \nonumber \\ |\alpha_{1j}|^2&=&\frac{1+\alpha}{4}-\sin^2\chi, \text{and} \;\;
    |\overline{\alpha_{1j}}|^2=\sin^2\chi-\frac{\alpha}{2} \nonumber,
\end{eqnarray}
for $j=1$ to $j=3$, and $0\le \chi \le 2\pi$. It is to be noted that although there are relative phases in each of the coefficients in $\ket{\psi_{o(e)}}$, the phases do not appear in the conditions~\eqref{dakhshini} and~\eqref{vrantibilas}, as the expectations are taken with respect to the $\sigma_z$ operator. \\
\section{How much precision can product probes provide?}
%%%%%%%%%%%%%%%%%%%%%%%apototo baad%%%%%%%%%%%%%%%%%%%%%%%%%
%-\textcolor{red}{Next we find the best precision obtainable with pure product probes, and the optimal product probe that provides such precision.}
%%%%%%%%%%%%%%%%%%%%%%%%%%%%%%%%%%%%%%
%%%%%%%%%%%%%%%%%%%%%%%%%%%%%%%%%%%%%%
%Here we  find the minimum value of the QFI-based figure of merit, $\Tr(\mathcal{F}_Q^{-1}\mathcal{W})$, when minimized over all pure product states, as a function of $N$.
%%%%%%%%%%%%%%%%%%%%%%%%%%%%%%%%%%%%%%
%%%%%%%%%%%%%%%%%%%%%%%%%%%%%%%%%%%%%%
%Our aim is to find the best precision that can be attained in the estimation of multiple independent local fields, each oriented along the Pauli-$z$ direction, 
%present in the Hamiltonian~\eqref{akashleena}, 
%while considering the weight matrix, $\mathcal{W}$, if the input probe used is a pure product one. 
\begin{theorem}
The lower bound of the QFI-based figure of merit, $\Tr(WF_Q^{-1})$, is given by $4N[(N-1)\alpha^2+1]$ for the encoding Hamiltonian, $H=\hbar \sum_{i=1}^N h_i \sigma_z^i$, and weight matrix, $\overline{W}$, if the optimization is performed over all pure product states. Here $\overline{W}=\sum_{i=1}^{N}a\ketbra{i}{i}+\sum_{i,j=1, i\ne j}^{N}b\ketbra{i}{j}$, where 
$a=16[(N-1)\alpha^2+1]$, and $b=16[(N-2)\alpha^2+2\alpha]$, for all $0<\alpha<1$ and a given value of $N$.
\end{theorem}
%\begin{theorem}
%The minimum value of the QFI-based figure of merit, in the estimation of $N$ independent single-body field terms corresponding to the encoding Hamiltonian, $H=\hbar \sum_{i=1}^N h_i \sigma_z^i$, and weight matrix, $\mathcal{W}$, where $\ket{\mathcal{W}}=\sum_{i=0}^{N-1} \ket{i}$ in the computational basis, is given by $N/4$, if the optimization is performed over all pure product states. 
%\textcolor{magenta}{In the estimation of $N$ independent local fields corresponding to the encoding Hamiltonian, $H=\hbar \sum_{i=1}^N h_i \sigma_z^i$, and weight matrix, $\mathcal{W}$, where $\ket{\mathcal{W}}=\sum_{i=0}^{N-1} \ket{i}$ in the computational basis, the minimum value of the quantity, $\Tr(\mathcal{W}\mathcal{F}_Q^{-1})$, is given by $N/4$, if the optimization is performed over all pure product states.}
%\end{theorem}
The value of the bound obtainable using the optimal pure product state is strictly larger than that using the optimal state, which is genuinely multipartite entangled.\\

\noindent \textit{\textbf{Proof.}} The input probe is considered to be a tensor product of $N$ single-qubit pure states, given by
\begin{eqnarray}
\ket{\psi_0^P}=\prod_{i=1}^N \left(\cos\frac{\theta_i}{2}\ket{0}+e^{i\phi_i}\sin\frac{\theta_i}{2}\ket{1}\right),
\end{eqnarray}
where $P$ in the superscript of $\ket{\psi_0^P}$ denotes that the state is product.  Our aim is to optimize over this class of states and find the minimum value of the quantum Fisher information-based figure of merit,  $\Tr(\overline{W}(\mathcal{F}_Q^P)^{-1})$, where $\mathcal{F}_Q^P$ is the QFIM corresponding to the state, $\ket{\psi_0^P}$, and encoding Hamiltonian, $H$.
%, corresponding to the state, $\ket{\psi_0^P}$. 
%From Eq.~\eqref{vrantibilas},
Now, we see that, for input probe $ \ket{\psi_0^P}$, the off-diagonal terms of $\mathcal{F}_Q^P$ are all equal to $0$, i.e. 
\begin{eqnarray}
(\mathcal{F}_Q^P)_{ij} &=& 4\Big[\bra{\psi_0^P} \sigma_z^i  \otimes \sigma_z^j  \ket{\psi_0^P} \nonumber \\ &-&   \bra{\psi_0^P} \sigma_z^i  \ket{\psi_0^P}\bra{\psi_0^P} \sigma_z^j \ket{\psi_0^P}\Big] 
= 0,
\end{eqnarray}
for $\forall i, j$ and $i \ne j$. Therefore, $\mathcal{F}_Q^P$, and hence $({\mathcal{F}_Q^P})^{-1}$, are diagonal in the computational basis, with the diagonal elements of $({\mathcal{F}_Q^P})^{-1}$ being given by $(({\mathcal{F}_Q^P})^{-1})_{ii}=1/[4(1-\Tr(\sigma_z^i \rho_i))^2]$, where $\rho_i=\Tr_{\bar{i}}\ketbra{\psi_0^P}{\psi_0^P}$. Further, since we choose a particular form of the weight matrix, $\overline{W}$, we obtain the quantum Fisher information-based figure of merit, $\mathcal{V}$, given by
\begin{eqnarray}
\label{megh_muluke}
\mathcal{V}=\Tr[\overline{W}({\mathcal{F}_Q^P})^{-1}]=\sum_{i=1}^N\frac{a}{4[1-\Tr(\sigma_z^i \rho_i)^2]},
\end{eqnarray}
where $a=16[(N-1)\alpha^2+1]$.
If we evaluate the quantity, $\Tr(\sigma_z^i \rho_i)$, corresponding to the state, $\ket{\psi_0^P}$, it provides a value equal to $\cos\theta_i$. Therefore we obtain
\begin{eqnarray}
\label{jonaki_der_bari}    \mathcal{V}=\sum_{i=1}^N\frac{a}{4\sin^2\theta_i},
\end{eqnarray}
where $a=16[(N-1)\alpha^2+1]$.
Now, we minimize the quantity, $\mathcal{V}$, with respect to $ \ket{\psi_0^P}$, i.e. minimize the right hand side of Eq.~\eqref{jonaki_der_bari} with respect to all $\theta_i$. Further, since all $\theta_i$ are independent, the minimization can be taken inside the summation, and each $\theta_i$ can be minimized independently, which essentially reduces the optimization to
\begin{eqnarray}
\min_{\ket{\psi_0^P}}\mathcal{V} =\sum_{i=1}^N\min_{\theta_i}\frac{a}{4\sin^2\theta_i}=\frac{Na}{4},
\end{eqnarray}
with $a=16[(N-1)\alpha^2+1]$, 
where the optimal value of $\theta_i=\theta_i^{opt}=\pi/2$, $\forall i$. Therefore it is proved that the optimal pure product state gives a minimum value of the quantum Fisher information-based figure of merit
%quantity, $\Tr(\mathcal{F}_Q^{-1}\mathcal{W})$, 
equal to $4N[(N-1)\alpha^2+1]$ for an $N$-party system. The value is larger than that obtained using GHZ state for $N\ge 2$. \hfill $\blacksquare$
\\

\begin{corollary}
The optimal pure product probe which saturates the lower bound of $\Tr(WF_Q^{-1})$ is given by $\otimes_{i=1}^N\left(\ket{0}+\exp(i\phi_i)\ket{1}\right)/\sqrt{2}$, where $\phi_i$ denotes the relative phase of the $i^{\text{th}}$ party
%. This result corresponds to the case where the 
corresponding to the Hamiltonian is $H=\hbar \sum_{i=1}^N h_i \sigma_z^i$, and weight matrix is $\overline{W}$. 
\end{corollary}

\section{Mixed states do not provide better precision}
%\label{chandra_prabha}
Here we ask the question whether the lower bound  of QFI-based figure of merit given in theorem~\eqref{bound_gen} is attainable by any mixed state for an arbitrary encoding Hamiltonian, $\widetilde{H}$, and an arbitrary positive semi-definite weight matrix, $W$. 
%%%%%%%%%%%%%%%%%%%%%%%%%%%%%%%%%%
%%%%%%%%%%%%%%%%%%%%%%%%%%%%%%%%%%
%We find that it is not, i.e. the best precision in the estimation of independent local fields is obtainable only by pure states, and any mixed state provides a precision lower than that. 
%%%%%%%%%%%%%%%%%%%%%%%%%%%%%%%%%
%%%%%%%%%%%%%%%%%%%%%%%%%%%%%%%%%

\begin{theorem}
\label{upalabdhi}
The minimum value of the QFI-based figure of merit, $\Tr(WF_Q^{-1})$, 
is not attainable by mixed states, in the estimation of $N$ independent field strengths of the encoding Hamiltonian,  $\widetilde{H}=\hbar \sum_{i=1}^N h_i H_i$, and an arbitrary positive semi-definite weight matrix, $W$.
%in the estimation of $N$ single-party independent field strengths of the $N$-party encoding Hamiltonian,  $\widetilde{H}=\hbar \sum_{i=1}^N h_i H_i$, containing only one-body terms, and an arbitrary weight matrix, $W$, cannot be attained by any mixed state, i.e. the minimum value obtained using mixed probe states is larger than that using pure probe states. 
\end{theorem}

\noindent \textit{Proof.}  Our aim is to prove that the equality in theorem 1 of the main text is achieved only by pure states, while for mixed states, the inequality holds with greater than sign. 
%We begin by considering the the inequality provided in~\eqref{gantabya}, which is true for arbitrary probe states, including mixed states, and arbitrary encodings, corresponding to an arbitrary choice of the weight matrix, $W$. 
Considering the inequation, $ \Tr(WF_Q^{-1})\ge \Tr(W^{1/2})^2/\Tr(F_Q)$, our aim is to minimize the quantity, $\Tr(W^{1/2})^2/\Tr(F_Q)$, over all arbitrary input states, including mixed ones, corresponding to the encoding Hamiltonian, $H$. 
So, we essentially have to find whether there exists any mixed state which maximizes the quantity, $\Tr(F_Q)$. Let the encoded state after evolving using the Hamiltonian, $H$, be given by $\rho_h=\sum_{i} \eta_h^i \ketbra{\eta_h^i}{\eta_h^i}$. 
%The subscript, $h$, means that the eigenvalues and eigenvectors of the encoded state depend on the set of parameters, $h=\{h_1,h_2,...,h_N\}$. 
The diagonal terms of the QFIM in this case are given by
\begin{eqnarray}
\label{batasia}
(F_Q^A)_{ii}=\sum_{\substack{n,n'\\ \eta_n+\eta_{n'}\ne 0}} 4\eta_n \frac{|\bra{\eta_n}\partial_i\rho_h\ket{\eta_{n'}}|^2}{(\eta_n+\eta_{n'})^2}, \;\; \forall i,
\end{eqnarray}
where the $A$ in the superscript in the left hand side denotes that this is the QFIM corresponding to any arbitrary state, which can be, in general mixed. Here $\partial_i$ denotes derivative with respect to the $i^{\text{th}}$ field strength, $h_i$. 
So we can write $\partial_i\rho_h=-i[H_i,\rho_h]$. From this, it follows that
\begin{eqnarray}
\bra{\eta_n}[H_i,\rho_h]\ket{\eta_{n'}} &=& 
\bra{\eta_n}[H_i-\langle H_i \rangle_{\rho_h} ,\rho_h]\ket{\eta_{n'}}\nonumber \\
&=& (\eta_{n'}-\eta_n) \bra{\eta_n}(H_i-\langle H_i \rangle_{\rho_h})\ket{\eta_{n'}}\nonumber
\end{eqnarray}
Now, using this in Eq.~\eqref{batasia}, we obtain, $\forall i$,
\begin{eqnarray}
\label{tasher_desh}
(F_Q^A)_{ii} &=& 4\sum_{\substack{n,n'\\ \eta_n+\eta_{n'}\ne 0}} \eta_n \frac{(\eta_{n'}-\eta_n)^2}{(\eta_{n'}+\eta_n)^2}|\bra{\eta_n}\Delta H_i \ket{\eta_{n'}}|^2 \nonumber \\
&\le& 
4\sum_{\substack{n,n'\\ \eta_n+\eta_{n'}\ne 0}} \eta_n r |\bra{\eta_n}\Delta H_i\ket{\eta_{n'}}|^2  \\
&=& 4 r  [\langle H_i^2\rangle_{\rho_h}-\langle H_i \rangle_{\rho_h}^2] , \nonumber
\end{eqnarray}
where $\Delta H_i= (H_i-\langle H_i \rangle_{\rho_h})$ and $r=(\eta_M-\eta_m)^2/(\eta_M+\eta_m)^2$. Here $\eta_{M(m)}$ is the maximum (minimum) eigenvalue of $\rho_h$. It has been used that the minimum value of the ratio of eigenvalues is $\min_{n,n'}(\eta_{n}/\eta_n')=\eta_m/\eta_M$.  Now we find the condition at which the inequality~\eqref{tasher_desh} becomes an equality, i.e. $(\eta_{n'}-\eta_n)^2/(\eta_{n'}+\eta_n)^2=r$, $\forall n, n'$ such that $\eta_n+\eta_{n'}\ne 0$. We see that the parameter, $r$, can take a maximum value equal to $1$. This means that corresponding to the maximum value of $r$, the quantity, $(\eta_{n'}-\eta_n)/(\eta_{n'}+\eta_n)=\pm 1$, which in turn implies either $\eta_n=0$ or $\eta_{n'}=0$, $\forall n, n'$ such that $\eta_n+\eta_{n'}\ne 0$. This suggests that if we consider the pair, $(\eta_n,\eta_{n'})$, one of them is always equal to $0$,  $\forall n, n'$. This can be possible if only one eigenvalue of $\rho_h$ is equal to $1$, and all others are $0$. It is therefore proved that $(\eta_{n'}-\eta_n)^2/(\eta_{n'}+\eta_n)^2=r_{max}=1$, only when the encoded state is pure. The quantity, $(F_Q^A)_{ii}$, is maximum at $r=1$, which implies that $\Tr(F_Q^A)$ is also maximum at $r=1$, which is possible only for a pure encoded state. We considered unitary encoding, and thus the input probe is also pure in such a scenario, which proves our claim in theorem~\ref{upalabdhi}.  \hfill $\blacksquare$ \\

Here we provide a short discussion after the proof of theorem~\ref{upalabdhi} in the main text.
If we consider $H_i=\sigma_z^i$ $\forall i$, 
%in inequation~\eqref{tasher_desh},
then we find that 
\begin{eqnarray}
\label{jyotsna}
    \max_{\rho_h} \Tr(\mathcal{F}_Q^A)=\max_{\rho_h}4r\left[N-\sum_i\Tr(\sigma_z^i \rho_h)^2\right] = 4N, \;\;\;
\end{eqnarray}
where $\mathcal{F}_Q^A$ is the  QFIM corresponding to the encoding Hamiltonian, $H$, and arbitrary input states, including mixed ones. The maximum in Eq.~\eqref{jyotsna} is attained only when the input probe is pure.  This is because we proved in theorem~\ref{upalabdhi} that $r=r_{max}=1$, only for a pure input probe. 
%Moreover, we have already seen in the preceding section that among the pure states, the GHZ state up to a relative phase, gives the optimal precision.The GHZ state up to a relative phase, minimizes the quantity, $4\left[N-\sum_i\Tr(\sigma_z^i \rho_h)^2\right]$, and gives a value $4N$. If we now assume the weight matrix to be $W=\mathcal{W}$, it then follows from Eq.~\eqref{jyotsna} that the minimum value of the quantity, $\Tr(\mathcal{W}^{1/2})^2/\Tr(\mathcal{F}_Q^A)$, in this case is $N/(4N)=1/4$, which is attainable only when the input probe is pure. 
Thus we find that the minimum value of the quantum Fisher information-based figure of merit is attainable only by pure probes, and mixed probes provide a precision worse than that using pure ones.\\

\section{Conclusion}
%\label{megher_chhaya}
It is known that
estimation of a common local field - common over several probes - can be  improved by employing entanglement as a resource. 
%
%The 
%However, a
Although the 
encoding process in this case utilizes a local Hamiltonian, there is an inherent “nonlocality” in the encoding process 
as a 
%in the form of 
a common local field is applied on all the probes. It may therefore seem
%Therefore it seems 
plausible that using independent local fields in the encoding process may
lead to the ineffectiveness of entanglement in the estimation. 
%We 
Here we 
%have 
proved that even if the encoding utilizes local independent fields, 
genuine multipartite entanglement is indispensable in the optimal probe, with the GHZ state in a certain orthonormal basis being a potential candidate. In particular, we find the parametric form of the most general optimal state for three parties.
%the Greenberger-Horne-Zeilinger state, in the eigenbasis of the Hamiltonian and up to a relative phase, is the only optimal probe.

We analyzed the estimation of multiple independent field  strengths of an arbitrary many-body encoding Hamiltonian comprising of one-body terms, and minimized 
%an attainable lower bound o
the estimation precision with respect to pure input probes. 
We provided an attainable bound on the precision in estimating multiple independent local fields in a quantum many-body system.
%which has been 
%used to gauge the accuracy of estimating the multiple field strengths. 
%Estimation of multiple parameters involve fundamental bounds which are given in terms of matrix inequalities. So to assess the actual precision in estimation, a figure of merit is defined in terms of a weight matrix, which follows certain properties that a suitable multiparameter 
%parameter 
%estimation measure should satisfy. We provided a lower bound of this figure of merit, in terms of an arbitrary weight matrix and an arbitrary but local encoding Hamiltonian comprising of an arbitrary number of parties. 
%Further, w
We considered a full-rank weight matrix, that takes into account both diagonal and off-diagonal elements of the quantum Fisher information matrix. This allowed us to consider not only the errors in the estimation of the different parameters, but also in some sense, the ``correlations'' or ``couplings'' between them. 
%Considering the field terms to be along the Pauli-$z$ direction, and for a particular choice of the weight matrix, we showed that the minimum value of the figure of merit is a constant, independent of the number of parties considered. Then we proved that this lower bound is attainable only by the Greenberger-Horne-Zeilinger state, in the eigenbasis of the Hamiltonian, if the optimization is performed over all pure states. 
Later we also showed that the lower bound that we provided is never attainable by any mixed state. We also provided the minimum value of the figure of merit that is obtainable using pure product probes.
%This is intriguing in the sense that even if we estimate multiple independent fields corresponding to a local Hamiltonian, it is necessary that the genuine-multiparty entangled GHZ state is used as the probe to attain 
%suffices to be optimal in attaining 
%the
%best metrological precision. 
%
%
%This can be intuitively understood as follows.
%The Hamiltonian is highly symmetric in each party, so it is apparent that the input probe should also be. Further, entangled probes are useful in single-parameter estimation, so genuinely entangled probes, that are symmetric in each party, are a natural choice in the multiparameter case. So the GHZ state appears to be a valid option. However, since we fix the Hamiltonian to be along the Pauli-$z$ direction, this breaks down the local unitary equivalence of the input probe, fixing only the GHZ state in the energy eigenbasis to be the optimal one.

The uncertainties of two non-commuting observables endorse quantumness into a system, with a prime example being the Heisenberg principle. Here we find that even if the parameters to be estimated are attached to 
%many-body Hamiltonian is 
``highly'' commuting 
%terms 
items - comprising of only single-body terms - there is a necessity of a genuine multiparty entangled state as the optimal probe. 

%We also provided the minimum value of the figure of merit that is obtainable using pure product probes. In fact, 
%We further proved that the difference between the figure of merit obtained using the optimal product state and the optimal state,  gradually increases with increasing number of parties. We also showed that the lower bound that we provided is never attainable by any mixed state.
%, therefore signifying that the GHZ state, up to a relative phase is the only choice as the optimal probe in the estimation of independent field strengths of arbitrary but local encoding Hamiltonians. 

En route, we proved that the choice of the identity operator as the weight matrix, and so ignoring the off-diagonal terms in the QFIM, does not necessitate the deployment of genuine multiparty entanglement in input probes for optimal estimation, and the optimal probe can be a pure product. 
%Intuitively t
Therefore, the choice of the weight matrix considered, involving both diagonal and off-diagonal elements, plays a crucial
%, and plays a significant 
role in necessitating 
%the role of 
genuine multiparty entangled probes as optimal ones.

\appendix
\section{Remark on theorem~\ref{bound_gen}}
\label{appen1}
It is to be noted that the maximum of $\Tr(F_Q)$ is obtained when $\ket{\psi_0}$ is pure product. However this, in general, is not the state which provides the minimum value of $\Tr(WF_Q^{-1})$. 
For every choice of probe state, the value of $\Tr(WF_Q^{-1})$ is always higher than or at most equal to $\Tr(W\widetilde{F}_Q^{-1})$. It happens that pure product states maximize the value of $\Tr(\widetilde{F}_Q)$, and corresponding to these states, the quantity, $\Tr(WF_Q^{-1})$ may have a value greater than or equal to $\Tr(W\widetilde{F}_Q^{-1})$. This certainly does not imply that $\Tr(WF_Q^{-1})$ is minimum for pure product states that maximize $\Tr(F_Q)$.
In fact, in the succeeding subsections, we will show that for a certain choice of the weight matrix, $W$, pure product probes never attain the lower bound of $\Tr(W\widetilde{F}_Q^{-1})$. \\

% In theorem 1 of the main text of this section, we consider the quantum Fisher information-based figure of merit, given by $\Tr(W F_Q^{-1})$, and minimize it with respect to all $N$-party pure input probes  for an arbitrary choice of $W$, and an arbitrary encoding many-body Hamiltonian, $\widetilde{H}$, comprising of a sum of single-body terms. Then we show that specifically  for  $W=\mathcal{W}$ and $F_Q=\mathcal{F}_Q$, the lower bound is a constant and  independent of the value of the number of parties. 
% Additionally, we find the condition of saturability of the lower bound that we provide.
% Further, in subsection~\ref{udashi}, still considering the weight matrix to be $\mathcal{W}$ and the QFIM to be $\mathcal{F}_Q$, we find that optimal precision is attainable using only the Greenberger-Horne-Zeilinger state, up to a relative phase, when the optimization is performed over all pure states. In subsection~\ref{chhobi_o_chhaya}, we find the minimum value of the quantum Fisher information-based figure of merit when the optimization is performed over all pure product states. We eventually show that the difference between the figures of merit obtained using the optimal product  and the GHZ states  gradually increases with an increasing number of parties. \\

\section{Calculation of square root of the weight matrix} 
\label{appen}
Let the square root of a $N\times N$ matrix, $W_0$, be given by 
\begin{eqnarray}
    W_0^{1/2} = u\sum_{i=1}^N\ketbra{i}{i} +  v\sum_{\substack{i,j=1\\i\ne j}}^N\ketbra{i}{j},
\end{eqnarray}
where $u>v$. Therefore the square of this matrix is given by
\begin{eqnarray}
    W_0 &=& \left(u\sum_{i=1}^N\ketbra{i}{i} +  v\sum_{\substack{i,j=1\\i\ne j}}^N\ketbra{i}{j}\right) \nonumber \\
   && \left(u\sum_{k=1}^N\ketbra{k}{k} +  v\sum_{\substack{k,l=1\\k\ne l}}^N\ketbra{k}{l}\right) \nonumber \\
   &=& u^2\sum_{i=1}^N\ketbra{i}{i} +  uv\sum_{\substack{i,l=1\\i\ne l}}^N\ketbra{i}{l} + uv\sum_{\substack{k,i=1\\k\ne i}}^N\ketbra{i}{k} \nonumber\\
   && + v^2 \sum_{\substack{i,j,k,l=1\\j\ne i\\l \ne k}}^N\ketbra{i}{l}.
\end{eqnarray}
Now the summation in the last term can be divided into two parts, viz. $i \ne l$, for which $j$ runs from $1$ to $N-2$, and $i = l$, for which $j$ runs from $1$ to $N-1$. This finally gives
\begin{eqnarray}
    W_0= u_1\sum_{i=1}^N\ketbra{i}{i} +  v_1\sum_{\substack{i,j=1\\i\ne j}}^N\ketbra{i}{j},
\end{eqnarray}
where $u_1=u^2+(N-1)v^2$ and $v_1=2uv+(N-2)v^2$, for a given value of $N$. For the matrix, $\overline{W}^{1/2}$, $u=4$ and $v=4\alpha$, where $0<\alpha<1$. Therefore in this scenario, $u_1=16[(N-1)\alpha^2+1]$ and $v_1=16[(N-2)\alpha^2+2\alpha]$.
\bibliography{h1_h2}

\begin{thebibliography}{59}%
\makeatletter
\providecommand \@ifxundefined [1]{%
 \@ifx{#1\undefined}
}%
\providecommand \@ifnum [1]{%
 \ifnum #1\expandafter \@firstoftwo
 \else \expandafter \@secondoftwo
 \fi
}%
\providecommand \@ifx [1]{%
 \ifx #1\expandafter \@firstoftwo
 \else \expandafter \@secondoftwo
 \fi
}%
\providecommand \natexlab [1]{#1}%
\providecommand \enquote  [1]{``#1''}%
\providecommand \bibnamefont  [1]{#1}%
\providecommand \bibfnamefont [1]{#1}%
\providecommand \citenamefont [1]{#1}%
\providecommand \href@noop [0]{\@secondoftwo}%
\providecommand \href [0]{\begingroup \@sanitize@url \@href}%
\providecommand \@href[1]{\@@startlink{#1}\@@href}%
\providecommand \@@href[1]{\endgroup#1\@@endlink}%
\providecommand \@sanitize@url [0]{\catcode `\\12\catcode `\$12\catcode `\&12\catcode `\#12\catcode `\^12\catcode `\_12\catcode `\%12\relax}%
\providecommand \@@startlink[1]{}%
\providecommand \@@endlink[0]{}%
\providecommand \url  [0]{\begingroup\@sanitize@url \@url }%
\providecommand \@url [1]{\endgroup\@href {#1}{\urlprefix }}%
\providecommand \urlprefix  [0]{URL }%
\providecommand \Eprint [0]{\href }%
\providecommand \doibase [0]{http://dx.doi.org/}%
\providecommand \selectlanguage [0]{\@gobble}%
\providecommand \bibinfo  [0]{\@secondoftwo}%
\providecommand \bibfield  [0]{\@secondoftwo}%
\providecommand \translation [1]{[#1]}%
\providecommand \BibitemOpen [0]{}%
\providecommand \bibitemStop [0]{}%
\providecommand \bibitemNoStop [0]{.\EOS\space}%
\providecommand \EOS [0]{\spacefactor3000\relax}%
\providecommand \BibitemShut  [1]{\csname bibitem#1\endcsname}%
\let\auto@bib@innerbib\@empty
%</preamble>
\bibitem [{\citenamefont {Wootters}(1981)}]{Wooters}%
  \BibitemOpen
  \bibfield  {author} {\bibinfo {author} {\bibfnamefont {W.~K.}\ \bibnamefont {Wootters}},\ }\bibfield  {title} {\enquote {\bibinfo {title} {Statistical distance and hilbert space},}\ }\href {\doibase 10.1103/PhysRevD.23.357} {\bibfield  {journal} {\bibinfo  {journal} {Phys. Rev. D}\ }\textbf {\bibinfo {volume} {23}},\ \bibinfo {pages} {357} (\bibinfo {year} {1981})}\BibitemShut {NoStop}%
\bibitem [{\citenamefont {Kay}(1993)}]{asymptotic}%
  \BibitemOpen
  \bibfield  {author} {\bibinfo {author} {\bibfnamefont {S.~M.}\ \bibnamefont {Kay}},\ }\bibfield  {title} {\enquote {\bibinfo {title} {Fundamentals of statistical signal processing: Estimation theory},}\ }\href@noop {} {\bibfield  {journal} {\bibinfo  {journal} {{Quantum}}\ } (\bibinfo {year} {1993})}\BibitemShut {NoStop}%
\bibitem [{\citenamefont {Braunstein}\ and\ \citenamefont {Caves}(1994)}]{Braunstein1}%
  \BibitemOpen
  \bibfield  {author} {\bibinfo {author} {\bibfnamefont {S.~L.}\ \bibnamefont {Braunstein}}\ and\ \bibinfo {author} {\bibfnamefont {C.~M.}\ \bibnamefont {Caves}},\ }\bibfield  {title} {\enquote {\bibinfo {title} {Statistical distance and the geometry of quantum states},}\ }\href {\doibase 10.1103/PhysRevLett.72.3439} {\bibfield  {journal} {\bibinfo  {journal} {Phys. Rev. Lett.}\ }\textbf {\bibinfo {volume} {72}},\ \bibinfo {pages} {3439} (\bibinfo {year} {1994})}\BibitemShut {NoStop}%
\bibitem [{\citenamefont {Matsumoto}(2002)}]{Matsumoto}%
  \BibitemOpen
  \bibfield  {author} {\bibinfo {author} {\bibfnamefont {K}~\bibnamefont {Matsumoto}},\ }\bibfield  {title} {\enquote {\bibinfo {title} {A new approach to the cramér-rao-type bound of the pure-state model},}\ }\href {\doibase 10.1088/0305-4470/35/13/307} {\bibfield  {journal} {\bibinfo  {journal} {Journal of Physics A: Mathematical and General}\ }\textbf {\bibinfo {volume} {35}},\ \bibinfo {pages} {3111} (\bibinfo {year} {2002})}\BibitemShut {NoStop}%
\bibitem [{\citenamefont {Giovannetti}\ \emph {et~al.}(2011)\citenamefont {Giovannetti}, \citenamefont {Lloyd},\ and\ \citenamefont {Maccone}}]{macconerev}%
  \BibitemOpen
  \bibfield  {author} {\bibinfo {author} {\bibfnamefont {V.}~\bibnamefont {Giovannetti}}, \bibinfo {author} {\bibfnamefont {S.}~\bibnamefont {Lloyd}}, \ and\ \bibinfo {author} {\bibfnamefont {L.}~\bibnamefont {Maccone}},\ }\bibfield  {title} {\enquote {\bibinfo {title} {Advances in quantum metrology},}\ }\href {\doibase 10.1038/nphoton.2011.35} {\bibfield  {journal} {\bibinfo  {journal} {Nat. Photonics}\ }\textbf {\bibinfo {volume} {5}},\ \bibinfo {pages} {222} (\bibinfo {year} {2011})}\BibitemShut {NoStop}%
\bibitem [{\citenamefont {Spagnolo}\ \emph {et~al.}(2012)\citenamefont {Spagnolo}, \citenamefont {Aparo}, \citenamefont {Vitelli}, \citenamefont {Crespi}, \citenamefont {Ramponi}, \citenamefont {Osellame}, \citenamefont {Mataloni},\ and\ \citenamefont {Sciarrino}}]{expt}%
  \BibitemOpen
  \bibfield  {author} {\bibinfo {author} {\bibfnamefont {N.}~\bibnamefont {Spagnolo}}, \bibinfo {author} {\bibfnamefont {L.}~\bibnamefont {Aparo}}, \bibinfo {author} {\bibfnamefont {C.}~\bibnamefont {Vitelli}}, \bibinfo {author} {\bibfnamefont {A.}~\bibnamefont {Crespi}}, \bibinfo {author} {\bibfnamefont {R.}~\bibnamefont {Ramponi}}, \bibinfo {author} {\bibfnamefont {R.}~\bibnamefont {Osellame}}, \bibinfo {author} {\bibfnamefont {P.}~\bibnamefont {Mataloni}}, \ and\ \bibinfo {author} {\bibfnamefont {F.}~\bibnamefont {Sciarrino}},\ }\bibfield  {title} {\enquote {\bibinfo {title} {Quantum interferometry with three-dimensional geometry},}\ }\href {\doibase https://doi.org/10.1038/srep00862} {\ \bibinfo {series} {Scientific Reports},\ \textbf {\bibinfo {volume} {2}} (\bibinfo {year} {2012}),\ https://doi.org/10.1038/srep00862}\BibitemShut {NoStop}%
\bibitem [{\citenamefont {Tóth}\ and\ \citenamefont {Apellaniz}(2014)}]{rev2014}%
  \BibitemOpen
  \bibfield  {author} {\bibinfo {author} {\bibfnamefont {G.}~\bibnamefont {Tóth}}\ and\ \bibinfo {author} {\bibfnamefont {I.}~\bibnamefont {Apellaniz}},\ }\bibfield  {title} {\enquote {\bibinfo {title} {Quantum metrology from a quantum information science perspective},}\ }\href {\doibase 10.1088/1751-8113/47/42/424006} {\bibfield  {journal} {\bibinfo  {journal} {Journal of Physics A: Mathematical and Theoretical}\ }\textbf {\bibinfo {volume} {47}},\ \bibinfo {pages} {424006} (\bibinfo {year} {2014})}\BibitemShut {NoStop}%
\bibitem [{\citenamefont {Demkowicz-Dobrzański}\ \emph {et~al.}(2015)\citenamefont {Demkowicz-Dobrzański}, \citenamefont {Jarzyna},\ and\ \citenamefont {Kołodyński}}]{rev2015}%
  \BibitemOpen
  \bibfield  {author} {\bibinfo {author} {\bibfnamefont {R.}~\bibnamefont {Demkowicz-Dobrzański}}, \bibinfo {author} {\bibfnamefont {M.}~\bibnamefont {Jarzyna}}, \ and\ \bibinfo {author} {\bibfnamefont {J.}~\bibnamefont {Kołodyński}},\ }\bibfield  {title} {\enquote {\bibinfo {title} {Chapter four - quantum limits in optical interferometry},}\ }\href {\doibase https://doi.org/10.1016/bs.po.2015.02.003} {\ \bibinfo {series} {Progress in Optics},\ \textbf {\bibinfo {volume} {60}},\ \bibinfo {pages} {345--435} (\bibinfo {year} {2015})}\BibitemShut {NoStop}%
\bibitem [{\citenamefont {Ragy}\ \emph {et~al.}(2016{\natexlab{a}})\citenamefont {Ragy}, \citenamefont {Jarzyna},\ and\ \citenamefont {Demkowicz-Dobrza\ifmmode~\acute{n}\else \'{n}\fi{}ski}}]{Rafal}%
  \BibitemOpen
  \bibfield  {author} {\bibinfo {author} {\bibfnamefont {S.}~\bibnamefont {Ragy}}, \bibinfo {author} {\bibfnamefont {M.}~\bibnamefont {Jarzyna}}, \ and\ \bibinfo {author} {\bibfnamefont {R.}~\bibnamefont {Demkowicz-Dobrza\ifmmode~\acute{n}\else \'{n}\fi{}ski}},\ }\bibfield  {title} {\enquote {\bibinfo {title} {Compatibility in multiparameter quantum metrology},}\ }\href {\doibase 10.1103/PhysRevA.94.052108} {\bibfield  {journal} {\bibinfo  {journal} {Phys. Rev. A}\ }\textbf {\bibinfo {volume} {94}},\ \bibinfo {pages} {052108} (\bibinfo {year} {2016}{\natexlab{a}})}\BibitemShut {NoStop}%
\bibitem [{\citenamefont {Vidrighin}\ \emph {et~al.}(2014)\citenamefont {Vidrighin}, \citenamefont {Donati}, \citenamefont {Genoni}, \citenamefont {Jin}, \citenamefont {Kolthammer}, \citenamefont {Kim}, \citenamefont {Datta}, \citenamefont {Barbieri},\ and\ \citenamefont {Walmsley}}]{ref2}%
  \BibitemOpen
  \bibfield  {author} {\bibinfo {author} {\bibfnamefont {M.~D.}\ \bibnamefont {Vidrighin}}, \bibinfo {author} {\bibfnamefont {G.}~\bibnamefont {Donati}}, \bibinfo {author} {\bibfnamefont {M.~G.}\ \bibnamefont {Genoni}}, \bibinfo {author} {\bibfnamefont {X.}~\bibnamefont {Jin}}, \bibinfo {author} {\bibfnamefont {W.~S.}\ \bibnamefont {Kolthammer}}, \bibinfo {author} {\bibfnamefont {M.S.}\ \bibnamefont {Kim}}, \bibinfo {author} {\bibfnamefont {A.}~\bibnamefont {Datta}}, \bibinfo {author} {\bibfnamefont {M.}~\bibnamefont {Barbieri}}, \ and\ \bibinfo {author} {\bibfnamefont {I.~A.}\ \bibnamefont {Walmsley}},\ }\bibfield  {title} {\enquote {\bibinfo {title} {Joint estimation of phase and phase diffusion for quantum metrology},}\ }\href {\doibase 10.1038/ncomms4532} {\bibfield  {journal} {\bibinfo  {journal} {Nat. Commun.}\ }\textbf {\bibinfo {volume} {5}},\ \bibinfo {pages} {3532} (\bibinfo {year} {2014})}\BibitemShut {NoStop}%
\bibitem [{\citenamefont {Ragy}\ \emph {et~al.}(2016{\natexlab{b}})\citenamefont {Ragy}, \citenamefont {Jarzyna},\ and\ \citenamefont {Demkowicz-Dobrza\ifmmode~\acute{n}\else \'{n}\fi{}ski}}]{ref6}%
  \BibitemOpen
  \bibfield  {author} {\bibinfo {author} {\bibfnamefont {S.}~\bibnamefont {Ragy}}, \bibinfo {author} {\bibfnamefont {M.}~\bibnamefont {Jarzyna}}, \ and\ \bibinfo {author} {\bibfnamefont {R.}~\bibnamefont {Demkowicz-Dobrza\ifmmode~\acute{n}\else \'{n}\fi{}ski}},\ }\bibfield  {title} {\enquote {\bibinfo {title} {Compatibility in multiparameter quantum metrology},}\ }\href {\doibase 10.1103/PhysRevA.94.052108} {\bibfield  {journal} {\bibinfo  {journal} {Phys. Rev. A}\ }\textbf {\bibinfo {volume} {94}},\ \bibinfo {pages} {052108} (\bibinfo {year} {2016}{\natexlab{b}})}\BibitemShut {NoStop}%
\bibitem [{\citenamefont {Pezz\`e}\ \emph {et~al.}(2017)\citenamefont {Pezz\`e}, \citenamefont {Ciampini}, \citenamefont {Spagnolo}, \citenamefont {Humphreys}, \citenamefont {Datta}, \citenamefont {Walmsley}, \citenamefont {Barbieri}, \citenamefont {Sciarrino},\ and\ \citenamefont {Smerzi}}]{ref1}%
  \BibitemOpen
  \bibfield  {author} {\bibinfo {author} {\bibfnamefont {L.}~\bibnamefont {Pezz\`e}}, \bibinfo {author} {\bibfnamefont {M.~A.}\ \bibnamefont {Ciampini}}, \bibinfo {author} {\bibfnamefont {N.}~\bibnamefont {Spagnolo}}, \bibinfo {author} {\bibfnamefont {P.~C.}\ \bibnamefont {Humphreys}}, \bibinfo {author} {\bibfnamefont {A.}~\bibnamefont {Datta}}, \bibinfo {author} {\bibfnamefont {I.~A.}\ \bibnamefont {Walmsley}}, \bibinfo {author} {\bibfnamefont {M.}~\bibnamefont {Barbieri}}, \bibinfo {author} {\bibfnamefont {F.}~\bibnamefont {Sciarrino}}, \ and\ \bibinfo {author} {\bibfnamefont {A.}~\bibnamefont {Smerzi}},\ }\bibfield  {title} {\enquote {\bibinfo {title} {Optimal measurements for simultaneous quantum estimation of multiple phases},}\ }\href {\doibase 10.1103/PhysRevLett.119.130504} {\bibfield  {journal} {\bibinfo  {journal} {Phys. Rev. Lett.}\ }\textbf {\bibinfo {volume} {119}},\ \bibinfo {pages} {130504} (\bibinfo {year} {2017})}\BibitemShut {NoStop}%
\bibitem [{\citenamefont {Yousefjani}\ \emph {et~al.}(2017)\citenamefont {Yousefjani}, \citenamefont {Nichols}, \citenamefont {Salimi},\ and\ \citenamefont {Adesso}}]{ref7}%
  \BibitemOpen
  \bibfield  {author} {\bibinfo {author} {\bibfnamefont {R.}~\bibnamefont {Yousefjani}}, \bibinfo {author} {\bibfnamefont {R.}~\bibnamefont {Nichols}}, \bibinfo {author} {\bibfnamefont {S.}~\bibnamefont {Salimi}}, \ and\ \bibinfo {author} {\bibfnamefont {G.}~\bibnamefont {Adesso}},\ }\bibfield  {title} {\enquote {\bibinfo {title} {Estimating phase with a random generator: Strategies and resources in multiparameter quantum metrology},}\ }\href {\doibase 10.1103/PhysRevA.95.062307} {\bibfield  {journal} {\bibinfo  {journal} {Phys. Rev. A}\ }\textbf {\bibinfo {volume} {95}},\ \bibinfo {pages} {062307} (\bibinfo {year} {2017})}\BibitemShut {NoStop}%
\bibitem [{\citenamefont {Beau}\ and\ \citenamefont {del Campo}(2017)}]{mixed}%
  \BibitemOpen
  \bibfield  {author} {\bibinfo {author} {\bibfnamefont {M.}~\bibnamefont {Beau}}\ and\ \bibinfo {author} {\bibfnamefont {A.}~\bibnamefont {del Campo}},\ }\bibfield  {title} {\enquote {\bibinfo {title} {Nonlinear quantum metrology of many-body open systems},}\ }\href {\doibase 10.1103/PhysRevLett.119.010403} {\bibfield  {journal} {\bibinfo  {journal} {Phys. Rev. Lett.}\ }\textbf {\bibinfo {volume} {119}},\ \bibinfo {pages} {010403} (\bibinfo {year} {2017})}\BibitemShut {NoStop}%
\bibitem [{\citenamefont {Gessner}\ \emph {et~al.}(2018{\natexlab{a}})\citenamefont {Gessner}, \citenamefont {Pezz\`e},\ and\ \citenamefont {Smerzi}}]{ref8}%
  \BibitemOpen
  \bibfield  {author} {\bibinfo {author} {\bibfnamefont {M.}~\bibnamefont {Gessner}}, \bibinfo {author} {\bibfnamefont {L.}~\bibnamefont {Pezz\`e}}, \ and\ \bibinfo {author} {\bibfnamefont {A.}~\bibnamefont {Smerzi}},\ }\bibfield  {title} {\enquote {\bibinfo {title} {Sensitivity bounds for multiparameter quantum metrology},}\ }\href {\doibase 10.1103/PhysRevLett.121.130503} {\bibfield  {journal} {\bibinfo  {journal} {Phys. Rev. Lett.}\ }\textbf {\bibinfo {volume} {121}},\ \bibinfo {pages} {130503} (\bibinfo {year} {2018}{\natexlab{a}})}\BibitemShut {NoStop}%
\bibitem [{\citenamefont {Zhuang}\ \emph {et~al.}(2018)\citenamefont {Zhuang}, \citenamefont {Huang},\ and\ \citenamefont {Lee}}]{ref10}%
  \BibitemOpen
  \bibfield  {author} {\bibinfo {author} {\bibfnamefont {M.}~\bibnamefont {Zhuang}}, \bibinfo {author} {\bibfnamefont {J.}~\bibnamefont {Huang}}, \ and\ \bibinfo {author} {\bibfnamefont {C.}~\bibnamefont {Lee}},\ }\bibfield  {title} {\enquote {\bibinfo {title} {Multiparameter estimation via an ensemble of spinor atoms},}\ }\href {\doibase 10.1103/PhysRevA.98.033603} {\bibfield  {journal} {\bibinfo  {journal} {Phys. Rev. A}\ }\textbf {\bibinfo {volume} {98}},\ \bibinfo {pages} {033603} (\bibinfo {year} {2018})}\BibitemShut {NoStop}%
\bibitem [{\citenamefont {Gessner}\ \emph {et~al.}(2018{\natexlab{b}})\citenamefont {Gessner}, \citenamefont {Pezz\`e},\ and\ \citenamefont {Smerzi}}]{ref3}%
  \BibitemOpen
  \bibfield  {author} {\bibinfo {author} {\bibfnamefont {M.}~\bibnamefont {Gessner}}, \bibinfo {author} {\bibfnamefont {L.}~\bibnamefont {Pezz\`e}}, \ and\ \bibinfo {author} {\bibfnamefont {A.}~\bibnamefont {Smerzi}},\ }\bibfield  {title} {\enquote {\bibinfo {title} {Sensitivity bounds for multiparameter quantum metrology},}\ }\href {\doibase 10.1103/PhysRevLett.121.130503} {\bibfield  {journal} {\bibinfo  {journal} {Phys. Rev. Lett.}\ }\textbf {\bibinfo {volume} {121}},\ \bibinfo {pages} {130503} (\bibinfo {year} {2018}{\natexlab{b}})}\BibitemShut {NoStop}%
\bibitem [{\citenamefont {Yang}\ \emph {et~al.}(2019)\citenamefont {Yang}, \citenamefont {Pang}, \citenamefont {Zhou},\ and\ \citenamefont {Jordan}}]{ref9}%
  \BibitemOpen
  \bibfield  {author} {\bibinfo {author} {\bibfnamefont {J.}~\bibnamefont {Yang}}, \bibinfo {author} {\bibfnamefont {S.}~\bibnamefont {Pang}}, \bibinfo {author} {\bibfnamefont {Y.}~\bibnamefont {Zhou}}, \ and\ \bibinfo {author} {\bibfnamefont {A.~N.}\ \bibnamefont {Jordan}},\ }\bibfield  {title} {\enquote {\bibinfo {title} {Optimal measurements for quantum multiparameter estimation with general states},}\ }\href {\doibase 10.1103/PhysRevA.100.032104} {\bibfield  {journal} {\bibinfo  {journal} {Phys. Rev. A}\ }\textbf {\bibinfo {volume} {100}},\ \bibinfo {pages} {032104} (\bibinfo {year} {2019})}\BibitemShut {NoStop}%
\bibitem [{\citenamefont {Tan}\ and\ \citenamefont {Jeong}(2019)}]{review1}%
  \BibitemOpen
  \bibfield  {author} {\bibinfo {author} {\bibfnamefont {K.~C.}\ \bibnamefont {Tan}}\ and\ \bibinfo {author} {\bibfnamefont {H.}~\bibnamefont {Jeong}},\ }\bibfield  {title} {\enquote {\bibinfo {title} {{Nonclassical light and metrological power: An introductory review}},}\ }\href {\doibase 10.1116/1.5126696} {\bibfield  {journal} {\bibinfo  {journal} {AVS Quantum Sci.}\ }\textbf {\bibinfo {volume} {1}},\ \bibinfo {pages} {014701} (\bibinfo {year} {2019})}\BibitemShut {NoStop}%
\bibitem [{\citenamefont {Liu}\ \emph {et~al.}(2019)\citenamefont {Liu}, \citenamefont {Yuan}, \citenamefont {Lu},\ and\ \citenamefont {Wang}}]{ref11}%
  \BibitemOpen
  \bibfield  {author} {\bibinfo {author} {\bibfnamefont {J.}~\bibnamefont {Liu}}, \bibinfo {author} {\bibfnamefont {H.}~\bibnamefont {Yuan}}, \bibinfo {author} {\bibfnamefont {X.}~\bibnamefont {Lu}}, \ and\ \bibinfo {author} {\bibfnamefont {X.}~\bibnamefont {Wang}},\ }\bibfield  {title} {\enquote {\bibinfo {title} {Quantum fisher information matrix and multiparameter estimation},}\ }\href {\doibase 10.1088/1751-8121/ab5d4d} {\bibfield  {journal} {\bibinfo  {journal} {Journal of Physics A: Mathematical and Theoretical}\ }\textbf {\bibinfo {volume} {53}},\ \bibinfo {pages} {023001} (\bibinfo {year} {2019})}\BibitemShut {NoStop}%
\bibitem [{\citenamefont {Goldberg}\ \emph {et~al.}(2020)\citenamefont {Goldberg}, \citenamefont {Gianani}, \citenamefont {Barbieri}, \citenamefont {Sciarrino}, \citenamefont {Steinberg},\ and\ \citenamefont {Spagnolo}}]{ref4}%
  \BibitemOpen
  \bibfield  {author} {\bibinfo {author} {\bibfnamefont {A.~Z.}\ \bibnamefont {Goldberg}}, \bibinfo {author} {\bibfnamefont {I.}~\bibnamefont {Gianani}}, \bibinfo {author} {\bibfnamefont {M.}~\bibnamefont {Barbieri}}, \bibinfo {author} {\bibfnamefont {F.}~\bibnamefont {Sciarrino}}, \bibinfo {author} {\bibfnamefont {A.~M.}\ \bibnamefont {Steinberg}}, \ and\ \bibinfo {author} {\bibfnamefont {N.}~\bibnamefont {Spagnolo}},\ }\bibfield  {title} {\enquote {\bibinfo {title} {Multiphase estimation without a reference mode},}\ }\href {\doibase 10.1103/PhysRevA.102.022230} {\bibfield  {journal} {\bibinfo  {journal} {Phys. Rev. A}\ }\textbf {\bibinfo {volume} {102}},\ \bibinfo {pages} {022230} (\bibinfo {year} {2020})}\BibitemShut {NoStop}%
\bibitem [{\citenamefont {Gross}\ and\ \citenamefont {Caves}(2020)}]{ref5}%
  \BibitemOpen
  \bibfield  {author} {\bibinfo {author} {\bibfnamefont {J.~A.}\ \bibnamefont {Gross}}\ and\ \bibinfo {author} {\bibfnamefont {C.~M.}\ \bibnamefont {Caves}},\ }\bibfield  {title} {\enquote {\bibinfo {title} {One from many: estimating a function of many parameters},}\ }\href {\doibase 10.1088/1751-8121/abb9ed} {\bibfield  {journal} {\bibinfo  {journal} {Journal of Physics A: Mathematical and Theoretical}\ }\textbf {\bibinfo {volume} {54}},\ \bibinfo {pages} {014001} (\bibinfo {year} {2020})}\BibitemShut {NoStop}%
\bibitem [{\citenamefont {Kull}\ \emph {et~al.}(2020)\citenamefont {Kull}, \citenamefont {Guérin},\ and\ \citenamefont {Verstraete}}]{funda_9}%
  \BibitemOpen
  \bibfield  {author} {\bibinfo {author} {\bibfnamefont {I.}~\bibnamefont {Kull}}, \bibinfo {author} {\bibfnamefont {P.~A.}\ \bibnamefont {Guérin}}, \ and\ \bibinfo {author} {\bibfnamefont {F.}~\bibnamefont {Verstraete}},\ }\bibfield  {title} {\enquote {\bibinfo {title} {Uncertainty and trade-offs in quantum multiparameter estimation},}\ }\href {\doibase 10.1088/1751-8121/ab7f67} {\bibfield  {journal} {\bibinfo  {journal} {Journal of Physics A: Mathematical and Theoretical}\ }\textbf {\bibinfo {volume} {53}},\ \bibinfo {pages} {244001} (\bibinfo {year} {2020})}\BibitemShut {NoStop}%
\bibitem [{\citenamefont {Lu}\ and\ \citenamefont {Wang}(2021)}]{ref12}%
  \BibitemOpen
  \bibfield  {author} {\bibinfo {author} {\bibfnamefont {X.}~\bibnamefont {Lu}}\ and\ \bibinfo {author} {\bibfnamefont {X.}~\bibnamefont {Wang}},\ }\bibfield  {title} {\enquote {\bibinfo {title} {Incorporating heisenberg's uncertainty principle into quantum multiparameter estimation},}\ }\href {\doibase 10.1103/PhysRevLett.126.120503} {\bibfield  {journal} {\bibinfo  {journal} {Phys. Rev. Lett.}\ }\textbf {\bibinfo {volume} {126}},\ \bibinfo {pages} {120503} (\bibinfo {year} {2021})}\BibitemShut {NoStop}%
\bibitem [{\citenamefont {Goldberg}\ \emph {et~al.}(2021)\citenamefont {Goldberg}, \citenamefont {S\'anchez-Soto},\ and\ \citenamefont {Ferretti}}]{ref13}%
  \BibitemOpen
  \bibfield  {author} {\bibinfo {author} {\bibfnamefont {A.~Z.}\ \bibnamefont {Goldberg}}, \bibinfo {author} {\bibfnamefont {L.~L.}\ \bibnamefont {S\'anchez-Soto}}, \ and\ \bibinfo {author} {\bibfnamefont {H.}~\bibnamefont {Ferretti}},\ }\bibfield  {title} {\enquote {\bibinfo {title} {Intrinsic sensitivity limits for multiparameter quantum metrology},}\ }\href {\doibase 10.1103/PhysRevLett.127.110501} {\bibfield  {journal} {\bibinfo  {journal} {Phys. Rev. Lett.}\ }\textbf {\bibinfo {volume} {127}},\ \bibinfo {pages} {110501} (\bibinfo {year} {2021})}\BibitemShut {NoStop}%
\bibitem [{\citenamefont {Yang}\ \emph {et~al.}(2022)\citenamefont {Yang}, \citenamefont {Ru}, \citenamefont {An}, \citenamefont {Wang}, \citenamefont {Wang}, \citenamefont {Zhang},\ and\ \citenamefont {Li}}]{ref14}%
  \BibitemOpen
  \bibfield  {author} {\bibinfo {author} {\bibfnamefont {Y.}~\bibnamefont {Yang}}, \bibinfo {author} {\bibfnamefont {S.}~\bibnamefont {Ru}}, \bibinfo {author} {\bibfnamefont {M.}~\bibnamefont {An}}, \bibinfo {author} {\bibfnamefont {Y.}~\bibnamefont {Wang}}, \bibinfo {author} {\bibfnamefont {F.}~\bibnamefont {Wang}}, \bibinfo {author} {\bibfnamefont {P.}~\bibnamefont {Zhang}}, \ and\ \bibinfo {author} {\bibfnamefont {F.}~\bibnamefont {Li}},\ }\bibfield  {title} {\enquote {\bibinfo {title} {Multiparameter simultaneous optimal estimation with an su(2) coding unitary evolution},}\ }\href {\doibase 10.1103/PhysRevA.105.022406} {\bibfield  {journal} {\bibinfo  {journal} {Phys. Rev. A}\ }\textbf {\bibinfo {volume} {105}},\ \bibinfo {pages} {022406} (\bibinfo {year} {2022})}\BibitemShut {NoStop}%
\bibitem [{\citenamefont {G\'orecki}\ and\ \citenamefont {Demkowicz-Dobrza\ifmmode~\acute{n}\else \'{n}\fi{}ski}(2022)}]{funda2}%
  \BibitemOpen
  \bibfield  {author} {\bibinfo {author} {\bibfnamefont {W.}~\bibnamefont {G\'orecki}}\ and\ \bibinfo {author} {\bibfnamefont {R.}~\bibnamefont {Demkowicz-Dobrza\ifmmode~\acute{n}\else \'{n}\fi{}ski}},\ }\bibfield  {title} {\enquote {\bibinfo {title} {Multiparameter quantum metrology in the heisenberg limit regime: Many-repetition scenario versus full optimization},}\ }\href {\doibase 10.1103/PhysRevA.106.012424} {\bibfield  {journal} {\bibinfo  {journal} {Phys. Rev. A}\ }\textbf {\bibinfo {volume} {106}},\ \bibinfo {pages} {012424} (\bibinfo {year} {2022})}\BibitemShut {NoStop}%
\bibitem [{\citenamefont {Miyazaki}\ and\ \citenamefont {Matsumoto}(2022)}]{funda14}%
  \BibitemOpen
  \bibfield  {author} {\bibinfo {author} {\bibfnamefont {J.}~\bibnamefont {Miyazaki}}\ and\ \bibinfo {author} {\bibfnamefont {K.}~\bibnamefont {Matsumoto}},\ }\bibfield  {title} {\enquote {\bibinfo {title} {Imaginarity-free quantum multiparameter estimation},}\ }\href {\doibase 10.22331/q-2022-03-10-665} {\bibfield  {journal} {\bibinfo  {journal} {{Quantum}}\ }\textbf {\bibinfo {volume} {6}},\ \bibinfo {pages} {665} (\bibinfo {year} {2022})}\BibitemShut {NoStop}%
\bibitem [{\citenamefont {Reilly}\ \emph {et~al.}(2023)\citenamefont {Reilly}, \citenamefont {Wilson}, \citenamefont {J\"ager}, \citenamefont {Wilson},\ and\ \citenamefont {Holland}}]{ref15}%
  \BibitemOpen
  \bibfield  {author} {\bibinfo {author} {\bibfnamefont {J.~T.}\ \bibnamefont {Reilly}}, \bibinfo {author} {\bibfnamefont {J.~D.}\ \bibnamefont {Wilson}}, \bibinfo {author} {\bibfnamefont {S.~B.}\ \bibnamefont {J\"ager}}, \bibinfo {author} {\bibfnamefont {C.}~\bibnamefont {Wilson}}, \ and\ \bibinfo {author} {\bibfnamefont {M.~J.}\ \bibnamefont {Holland}},\ }\bibfield  {title} {\enquote {\bibinfo {title} {Optimal generators for quantum sensing},}\ }\href {\doibase 10.1103/PhysRevLett.131.150802} {\bibfield  {journal} {\bibinfo  {journal} {Phys. Rev. Lett.}\ }\textbf {\bibinfo {volume} {131}},\ \bibinfo {pages} {150802} (\bibinfo {year} {2023})}\BibitemShut {NoStop}%
\bibitem [{\citenamefont {M.~Szczykulska}\ and\ \citenamefont {Datta}(2016)}]{rev0}%
  \BibitemOpen
  \bibfield  {author} {\bibinfo {author} {\bibfnamefont {T.~Baumgratz}\ \bibnamefont {M.~Szczykulska}}\ and\ \bibinfo {author} {\bibfnamefont {A.}~\bibnamefont {Datta}},\ }\bibfield  {title} {\enquote {\bibinfo {title} {Multi-parameter quantum metrology},}\ }\href {\doibase 10.1080/23746149.2016.1230476} {\bibfield  {journal} {\bibinfo  {journal} {Advances in Physics: X}\ }\textbf {\bibinfo {volume} {1}},\ \bibinfo {pages} {621--639} (\bibinfo {year} {2016})},\ \Eprint {http://arxiv.org/abs/https://doi.org/10.1080/23746149.2016.1230476} {https://doi.org/10.1080/23746149.2016.1230476} \BibitemShut {NoStop}%
\bibitem [{\citenamefont {Hamann}\ \emph {et~al.}(2023)\citenamefont {Hamann}, \citenamefont {Sekatski},\ and\ \citenamefont {Dür}}]{dur}%
  \BibitemOpen
  \bibfield  {author} {\bibinfo {author} {\bibfnamefont {A.}~\bibnamefont {Hamann}}, \bibinfo {author} {\bibfnamefont {P.}~\bibnamefont {Sekatski}}, \ and\ \bibinfo {author} {\bibfnamefont {W.}~\bibnamefont {Dür}},\ }\href@noop {} {\enquote {\bibinfo {title} {Optimal distributed multiparameter estimation in noisy environments},}\ } (\bibinfo {year} {2023}),\ \Eprint {http://arxiv.org/abs/2306.01077} {arXiv:2306.01077 [quant-ph]} \BibitemShut {NoStop}%
\bibitem [{\citenamefont {Bhattacharyya}\ \emph {et~al.}(2024{\natexlab{a}})\citenamefont {Bhattacharyya}, \citenamefont {Ghoshal},\ and\ \citenamefont {Sen}}]{disorder}%
  \BibitemOpen
  \bibfield  {author} {\bibinfo {author} {\bibfnamefont {A.}~\bibnamefont {Bhattacharyya}}, \bibinfo {author} {\bibfnamefont {A.}~\bibnamefont {Ghoshal}}, \ and\ \bibinfo {author} {\bibfnamefont {U.}~\bibnamefont {Sen}},\ }\bibfield  {title} {\enquote {\bibinfo {title} {Enhancing precision of atomic clocks by tuning disorder in accessories},}\ }\href {\doibase 10.1103/PhysRevA.110.012620} {\bibfield  {journal} {\bibinfo  {journal} {Phys. Rev. A}\ }\textbf {\bibinfo {volume} {110}},\ \bibinfo {pages} {012620} (\bibinfo {year} {2024}{\natexlab{a}})}\BibitemShut {NoStop}%
\bibitem [{\citenamefont {Hamann}\ \emph {et~al.}(2024)\citenamefont {Hamann}, \citenamefont {Sekatski},\ and\ \citenamefont {Dür}}]{dur2}%
  \BibitemOpen
  \bibfield  {author} {\bibinfo {author} {\bibfnamefont {A.}~\bibnamefont {Hamann}}, \bibinfo {author} {\bibfnamefont {P.}~\bibnamefont {Sekatski}}, \ and\ \bibinfo {author} {\bibfnamefont {W.}~\bibnamefont {Dür}},\ }\bibfield  {title} {\enquote {\bibinfo {title} {Optimal distributed multi-parameter estimation in noisy environments},}\ }\href {\doibase 10.1088/2058-9565/ad37d5} {\bibfield  {journal} {\bibinfo  {journal} {Quantum Science and Technology}\ }\textbf {\bibinfo {volume} {9}},\ \bibinfo {pages} {035005} (\bibinfo {year} {2024})}\BibitemShut {NoStop}%
\bibitem [{\citenamefont {Bhattacharyya}\ \emph {et~al.}(2024{\natexlab{b}})\citenamefont {Bhattacharyya}, \citenamefont {Saha},\ and\ \citenamefont {Sen}}]{evenodd}%
  \BibitemOpen
  \bibfield  {author} {\bibinfo {author} {\bibfnamefont {A.}~\bibnamefont {Bhattacharyya}}, \bibinfo {author} {\bibfnamefont {D.}~\bibnamefont {Saha}}, \ and\ \bibinfo {author} {\bibfnamefont {U.}~\bibnamefont {Sen}},\ }\bibfield  {title} {\enquote {\bibinfo {title} {Quantum sensing of even- versus odd-body interactions},}\ }\href {https://arxiv.org/abs/2401.06729} {\  (\bibinfo {year} {2024}{\natexlab{b}})},\ \Eprint {http://arxiv.org/abs/2401.06729} {arXiv:2401.06729} \BibitemShut {NoStop}%
\bibitem [{\citenamefont {Jayakumar}\ \emph {et~al.}(2024)\citenamefont {Jayakumar}, \citenamefont {Mycroft}, \citenamefont {Barbieri},\ and\ \citenamefont {Stobińska}}]{Jayanth}%
  \BibitemOpen
  \bibfield  {author} {\bibinfo {author} {\bibfnamefont {J.}~\bibnamefont {Jayakumar}}, \bibinfo {author} {\bibfnamefont {M.~E.}\ \bibnamefont {Mycroft}}, \bibinfo {author} {\bibfnamefont {M.}~\bibnamefont {Barbieri}}, \ and\ \bibinfo {author} {\bibfnamefont {M.}~\bibnamefont {Stobińska}},\ }\bibfield  {title} {\enquote {\bibinfo {title} {Quantum-enhanced joint estimation of phase and phase diffusion},}\ }\href {\doibase 10.1088/1367-2630/ad5eb0} {\bibfield  {journal} {\bibinfo  {journal} {New Journal of Physics}\ }\textbf {\bibinfo {volume} {26}},\ \bibinfo {pages} {073016} (\bibinfo {year} {2024})}\BibitemShut {NoStop}%
\bibitem [{\citenamefont {Skotiniotis}\ \emph {et~al.}(2015)\citenamefont {Skotiniotis}, \citenamefont {Sekatski},\ and\ \citenamefont {Dür}}]{common_field6}%
  \BibitemOpen
  \bibfield  {author} {\bibinfo {author} {\bibfnamefont {M.}~\bibnamefont {Skotiniotis}}, \bibinfo {author} {\bibfnamefont {P.}~\bibnamefont {Sekatski}}, \ and\ \bibinfo {author} {\bibfnamefont {W.}~\bibnamefont {Dür}},\ }\bibfield  {title} {\enquote {\bibinfo {title} {Quantum metrology for the ising hamiltonian with transverse magnetic field},}\ }\href {\doibase 10.1088/1367-2630/17/7/073032} {\bibfield  {journal} {\bibinfo  {journal} {New Journal of Physics}\ }\textbf {\bibinfo {volume} {17}},\ \bibinfo {pages} {073032} (\bibinfo {year} {2015})}\BibitemShut {NoStop}%
\bibitem [{\citenamefont {Ouyang}\ and\ \citenamefont {Rengaswamy}(2023)}]{common_field5}%
  \BibitemOpen
  \bibfield  {author} {\bibinfo {author} {\bibfnamefont {Y.}~\bibnamefont {Ouyang}}\ and\ \bibinfo {author} {\bibfnamefont {N.}~\bibnamefont {Rengaswamy}},\ }\bibfield  {title} {\enquote {\bibinfo {title} {Describing quantum metrology with erasure errors using weight distributions of classical codes},}\ }\href {\doibase 10.1103/PhysRevA.107.022620} {\bibfield  {journal} {\bibinfo  {journal} {Phys. Rev. A}\ }\textbf {\bibinfo {volume} {107}},\ \bibinfo {pages} {022620} (\bibinfo {year} {2023})}\BibitemShut {NoStop}%
\bibitem [{\citenamefont {Deng}\ \emph {et~al.}(2021)\citenamefont {Deng}, \citenamefont {Chen}, \citenamefont {Zhang}, \citenamefont {Xu}, \citenamefont {Liu}, \citenamefont {Gao}, \citenamefont {Duan}, \citenamefont {Zhou}, \citenamefont {Cao},\ and\ \citenamefont {Hu}}]{common_field4}%
  \BibitemOpen
  \bibfield  {author} {\bibinfo {author} {\bibfnamefont {X.}~\bibnamefont {Deng}}, \bibinfo {author} {\bibfnamefont {S.}~\bibnamefont {Chen}}, \bibinfo {author} {\bibfnamefont {M.}~\bibnamefont {Zhang}}, \bibinfo {author} {\bibfnamefont {X.}~\bibnamefont {Xu}}, \bibinfo {author} {\bibfnamefont {J.}~\bibnamefont {Liu}}, \bibinfo {author} {\bibfnamefont {Z.}~\bibnamefont {Gao}}, \bibinfo {author} {\bibfnamefont {X.}~\bibnamefont {Duan}}, \bibinfo {author} {\bibfnamefont {M.}~\bibnamefont {Zhou}}, \bibinfo {author} {\bibfnamefont {L.}~\bibnamefont {Cao}}, \ and\ \bibinfo {author} {\bibfnamefont {Z.}~\bibnamefont {Hu}},\ }\bibfield  {title} {\enquote {\bibinfo {title} {Quantum metrology with precision reaching beyond $1/n$-scaling through $n$-probe entanglement-generating interactions},}\ }\href {\doibase 10.1103/PhysRevA.104.012607} {\bibfield  {journal} {\bibinfo  {journal} {Phys. Rev. A}\ }\textbf {\bibinfo {volume} {104}},\ \bibinfo {pages} {012607} (\bibinfo {year} {2021})}\BibitemShut {NoStop}%
\bibitem [{\citenamefont {Bhattacharyya}\ \emph {et~al.}(2024{\natexlab{c}})\citenamefont {Bhattacharyya}, \citenamefont {Ghoshal},\ and\ \citenamefont {Sen}}]{common_field3}%
  \BibitemOpen
  \bibfield  {author} {\bibinfo {author} {\bibfnamefont {A.}~\bibnamefont {Bhattacharyya}}, \bibinfo {author} {\bibfnamefont {A.}~\bibnamefont {Ghoshal}}, \ and\ \bibinfo {author} {\bibfnamefont {U.}~\bibnamefont {Sen}},\ }\bibfield  {title} {\enquote {\bibinfo {title} {Restoring metrological quantum advantage of measurement precision in a noisy scenario},}\ }\href {\doibase 10.1103/PhysRevA.109.052626} {\bibfield  {journal} {\bibinfo  {journal} {Phys. Rev. A}\ }\textbf {\bibinfo {volume} {109}},\ \bibinfo {pages} {052626} (\bibinfo {year} {2024}{\natexlab{c}})}\BibitemShut {NoStop}%
\bibitem [{\citenamefont {Salvia}\ \emph {et~al.}(2023)\citenamefont {Salvia}, \citenamefont {Mehboudi},\ and\ \citenamefont {Perarnau-Llobet}}]{common_field2}%
  \BibitemOpen
  \bibfield  {author} {\bibinfo {author} {\bibfnamefont {R.}~\bibnamefont {Salvia}}, \bibinfo {author} {\bibfnamefont {M.}~\bibnamefont {Mehboudi}}, \ and\ \bibinfo {author} {\bibfnamefont {M.}~\bibnamefont {Perarnau-Llobet}},\ }\bibfield  {title} {\enquote {\bibinfo {title} {Critical quantum metrology assisted by real-time feedback control},}\ }\href {\doibase 10.1103/PhysRevLett.130.240803} {\bibfield  {journal} {\bibinfo  {journal} {Phys. Rev. Lett.}\ }\textbf {\bibinfo {volume} {130}},\ \bibinfo {pages} {240803} (\bibinfo {year} {2023})}\BibitemShut {NoStop}%
\bibitem [{\citenamefont {Su}\ \emph {et~al.}(2024)\citenamefont {Su}, \citenamefont {Lu},\ and\ \citenamefont {Shi}}]{common_field1}%
  \BibitemOpen
  \bibfield  {author} {\bibinfo {author} {\bibfnamefont {Y.}~\bibnamefont {Su}}, \bibinfo {author} {\bibfnamefont {W.}~\bibnamefont {Lu}}, \ and\ \bibinfo {author} {\bibfnamefont {H.}~\bibnamefont {Shi}},\ }\bibfield  {title} {\enquote {\bibinfo {title} {Quantum metrology enhanced by the $xy$ spin interaction in a generalized tavis-cummings model},}\ }\href {\doibase 10.1103/PhysRevA.109.042614} {\bibfield  {journal} {\bibinfo  {journal} {Phys. Rev. A}\ }\textbf {\bibinfo {volume} {109}},\ \bibinfo {pages} {042614} (\bibinfo {year} {2024})}\BibitemShut {NoStop}%
\bibitem [{\citenamefont {Baumgratz}\ and\ \citenamefont {Datta}(2016)}]{mult_field1}%
  \BibitemOpen
  \bibfield  {author} {\bibinfo {author} {\bibfnamefont {T.}~\bibnamefont {Baumgratz}}\ and\ \bibinfo {author} {\bibfnamefont {A.}~\bibnamefont {Datta}},\ }\bibfield  {title} {\enquote {\bibinfo {title} {Quantum enhanced estimation of a multidimensional field},}\ }\href {\doibase 10.1103/PhysRevLett.116.030801} {\bibfield  {journal} {\bibinfo  {journal} {Phys. Rev. Lett.}\ }\textbf {\bibinfo {volume} {116}},\ \bibinfo {pages} {030801} (\bibinfo {year} {2016})}\BibitemShut {NoStop}%
\bibitem [{\citenamefont {Ho}\ \emph {et~al.}(2020)\citenamefont {Ho}, \citenamefont {Hakoshima}, \citenamefont {Matsuzaki}, \citenamefont {Matsuzaki},\ and\ \citenamefont {Kondo}}]{mult_field_noise1}%
  \BibitemOpen
  \bibfield  {author} {\bibinfo {author} {\bibfnamefont {L.}~\bibnamefont {Ho}}, \bibinfo {author} {\bibfnamefont {H.}~\bibnamefont {Hakoshima}}, \bibinfo {author} {\bibfnamefont {Y.}~\bibnamefont {Matsuzaki}}, \bibinfo {author} {\bibfnamefont {M.}~\bibnamefont {Matsuzaki}}, \ and\ \bibinfo {author} {\bibfnamefont {Y.}~\bibnamefont {Kondo}},\ }\bibfield  {title} {\enquote {\bibinfo {title} {Multiparameter quantum estimation under dephasing noise},}\ }\href {\doibase 10.1103/PhysRevA.102.022602} {\bibfield  {journal} {\bibinfo  {journal} {Phys. Rev. A}\ }\textbf {\bibinfo {volume} {102}},\ \bibinfo {pages} {022602} (\bibinfo {year} {2020})}\BibitemShut {NoStop}%
\bibitem [{\citenamefont {Le}\ \emph {et~al.}(2023)\citenamefont {Le}, \citenamefont {Nguyen},\ and\ \citenamefont {Ho}}]{mult_field_noise2}%
  \BibitemOpen
  \bibfield  {author} {\bibinfo {author} {\bibfnamefont {T.~K.}\ \bibnamefont {Le}}, \bibinfo {author} {\bibfnamefont {H.~Q.}\ \bibnamefont {Nguyen}}, \ and\ \bibinfo {author} {\bibfnamefont {L.~B.}\ \bibnamefont {Ho}},\ }\bibfield  {title} {\enquote {\bibinfo {title} {Variational quantum metrology for multiparameter estimation under dephasing noise},}\ }\href {https://arxiv.org/abs/2305.08289} {\  (\bibinfo {year} {2023})},\ \Eprint {http://arxiv.org/abs/2305.08289} {arXiv:2305.08289} \BibitemShut {NoStop}%
\bibitem [{\citenamefont {Giovannetti}\ \emph {et~al.}(2006)\citenamefont {Giovannetti}, \citenamefont {Lloyd},\ and\ \citenamefont {Maccone}}]{maccone}%
  \BibitemOpen
  \bibfield  {author} {\bibinfo {author} {\bibfnamefont {V.}~\bibnamefont {Giovannetti}}, \bibinfo {author} {\bibfnamefont {S.}~\bibnamefont {Lloyd}}, \ and\ \bibinfo {author} {\bibfnamefont {L.}~\bibnamefont {Maccone}},\ }\bibfield  {title} {\enquote {\bibinfo {title} {Quantum metrology},}\ }\href {\doibase 10.1103/PhysRevLett.96.010401} {\bibfield  {journal} {\bibinfo  {journal} {Phys. Rev. Lett.}\ }\textbf {\bibinfo {volume} {96}},\ \bibinfo {pages} {010401} (\bibinfo {year} {2006})}\BibitemShut {NoStop}%
\bibitem [{\citenamefont {Walther}\ \emph {et~al.}(2005)\citenamefont {Walther}, \citenamefont {Resch}, \citenamefont {Rudolph}, \citenamefont {Schenck}, \citenamefont {Weinfurter}, \citenamefont {Vedral}, \citenamefont {Aspelmeyer},\ and\ \citenamefont {Zeilinger}}]{state1}%
  \BibitemOpen
  \bibfield  {author} {\bibinfo {author} {\bibfnamefont {P.}~\bibnamefont {Walther}}, \bibinfo {author} {\bibfnamefont {K.~J.}\ \bibnamefont {Resch}}, \bibinfo {author} {\bibfnamefont {T.}~\bibnamefont {Rudolph}}, \bibinfo {author} {\bibfnamefont {E.}~\bibnamefont {Schenck}}, \bibinfo {author} {\bibfnamefont {H.}~\bibnamefont {Weinfurter}}, \bibinfo {author} {\bibfnamefont {V.}~\bibnamefont {Vedral}}, \bibinfo {author} {\bibfnamefont {M.}~\bibnamefont {Aspelmeyer}}, \ and\ \bibinfo {author} {\bibfnamefont {A.}~\bibnamefont {Zeilinger}},\ }\bibfield  {title} {\enquote {\bibinfo {title} {Experimental one-way quantum computing},}\ }\href {\doibase 10.1038/nature03347} {\bibfield  {journal} {\bibinfo  {journal} {Nature}\ }\textbf {\bibinfo {volume} {434}},\ \bibinfo {pages} {169--176} (\bibinfo {year} {2005})}\BibitemShut {NoStop}%
\bibitem [{\citenamefont {Su}\ \emph {et~al.}(2012)\citenamefont {Su}, \citenamefont {Zhao}, \citenamefont {Hao}, \citenamefont {Jia}, \citenamefont {Xie},\ and\ \citenamefont {Peng}}]{state2}%
  \BibitemOpen
  \bibfield  {author} {\bibinfo {author} {\bibfnamefont {X.}~\bibnamefont {Su}}, \bibinfo {author} {\bibfnamefont {Y.}~\bibnamefont {Zhao}}, \bibinfo {author} {\bibfnamefont {S.}~\bibnamefont {Hao}}, \bibinfo {author} {\bibfnamefont {X.}~\bibnamefont {Jia}}, \bibinfo {author} {\bibfnamefont {C.}~\bibnamefont {Xie}}, \ and\ \bibinfo {author} {\bibfnamefont {K.}~\bibnamefont {Peng}},\ }\bibfield  {title} {\enquote {\bibinfo {title} {Experimental preparation of eight-partite cluster state for photonic qumodes},}\ }\href {\doibase 10.1364/OL.37.005178} {\bibfield  {journal} {\bibinfo  {journal} {Opt. Lett.}\ }\textbf {\bibinfo {volume} {37}},\ \bibinfo {pages} {5178--5180} (\bibinfo {year} {2012})}\BibitemShut {NoStop}%
\bibitem [{\citenamefont {Ju}\ \emph {et~al.}(2021)\citenamefont {Ju}, \citenamefont {Yang},\ and\ \citenamefont {Xue}}]{state3}%
  \BibitemOpen
  \bibfield  {author} {\bibinfo {author} {\bibfnamefont {L.}~\bibnamefont {Ju}}, \bibinfo {author} {\bibfnamefont {M.}~\bibnamefont {Yang}}, \ and\ \bibinfo {author} {\bibfnamefont {P.}~\bibnamefont {Xue}},\ }\bibfield  {title} {\enquote {\bibinfo {title} {A proposal for preparation of cluster states with linear optics*},}\ }\href {\doibase 10.1088/1674-1056/abd74b} {\bibfield  {journal} {\bibinfo  {journal} {Chinese Physics B}\ }\textbf {\bibinfo {volume} {30}},\ \bibinfo {pages} {030306} (\bibinfo {year} {2021})}\BibitemShut {NoStop}%
\bibitem [{\citenamefont {Walther}\ \emph {et~al.}(2006)\citenamefont {Walther}, \citenamefont {Resch}, \citenamefont {Brukner},\ and\ \citenamefont {Zeilinger}}]{meas1}%
  \BibitemOpen
  \bibfield  {author} {\bibinfo {author} {\bibfnamefont {P.}~\bibnamefont {Walther}}, \bibinfo {author} {\bibfnamefont {K.~J.}\ \bibnamefont {Resch}}, \bibinfo {author} {\bibfnamefont {\ifmmode \check{C}\else~\v{C}\fi{}aslav}\ \bibnamefont {Brukner}}, \ and\ \bibinfo {author} {\bibfnamefont {A.}~\bibnamefont {Zeilinger}},\ }\bibfield  {title} {\enquote {\bibinfo {title} {Experimental entangled entanglement},}\ }\href {\doibase 10.1103/PhysRevLett.97.020501} {\bibfield  {journal} {\bibinfo  {journal} {Phys. Rev. Lett.}\ }\textbf {\bibinfo {volume} {97}},\ \bibinfo {pages} {020501} (\bibinfo {year} {2006})}\BibitemShut {NoStop}%
\bibitem [{\citenamefont {Zhang}\ \emph {et~al.}(2020)\citenamefont {Zhang}, \citenamefont {Zhang}, \citenamefont {Chen}, \citenamefont {Peng}, \citenamefont {Xu}, \citenamefont {Yin}, \citenamefont {Yu}, \citenamefont {Ye}, \citenamefont {Han}, \citenamefont {Xu}, \citenamefont {Chen}, \citenamefont {Li},\ and\ \citenamefont {Guo}}]{meas2}%
  \BibitemOpen
  \bibfield  {author} {\bibinfo {author} {\bibfnamefont {W.}~\bibnamefont {Zhang}}, \bibinfo {author} {\bibfnamefont {C.}~\bibnamefont {Zhang}}, \bibinfo {author} {\bibfnamefont {Z.}~\bibnamefont {Chen}}, \bibinfo {author} {\bibfnamefont {X.}~\bibnamefont {Peng}}, \bibinfo {author} {\bibfnamefont {X.}~\bibnamefont {Xu}}, \bibinfo {author} {\bibfnamefont {P.}~\bibnamefont {Yin}}, \bibinfo {author} {\bibfnamefont {S.}~\bibnamefont {Yu}}, \bibinfo {author} {\bibfnamefont {X.}~\bibnamefont {Ye}}, \bibinfo {author} {\bibfnamefont {Y.}~\bibnamefont {Han}}, \bibinfo {author} {\bibfnamefont {J.}~\bibnamefont {Xu}}, \bibinfo {author} {\bibfnamefont {G.}~\bibnamefont {Chen}}, \bibinfo {author} {\bibfnamefont {C.}~\bibnamefont {Li}}, \ and\ \bibinfo {author} {\bibfnamefont {G.}~\bibnamefont {Guo}},\ }\bibfield  {title} {\enquote {\bibinfo {title} {Experimental optimal verification of entangled states using local measurements},}\ }\href {\doibase 10.1103/PhysRevLett.125.030506} {\bibfield  {journal} {\bibinfo  {journal}
  {Phys. Rev. Lett.}\ }\textbf {\bibinfo {volume} {125}},\ \bibinfo {pages} {030506} (\bibinfo {year} {2020})}\BibitemShut {NoStop}%
\bibitem [{\citenamefont {Zhu}\ \emph {et~al.}(2022)\citenamefont {Zhu}, \citenamefont {Zhang}, \citenamefont {Wang}, \citenamefont {Xiao},\ and\ \citenamefont {Xue}}]{meas3}%
  \BibitemOpen
  \bibfield  {author} {\bibinfo {author} {\bibfnamefont {G.}~\bibnamefont {Zhu}}, \bibinfo {author} {\bibfnamefont {C.}~\bibnamefont {Zhang}}, \bibinfo {author} {\bibfnamefont {K.}~\bibnamefont {Wang}}, \bibinfo {author} {\bibfnamefont {L.}~\bibnamefont {Xiao}}, \ and\ \bibinfo {author} {\bibfnamefont {P.}~\bibnamefont {Xue}},\ }\bibfield  {title} {\enquote {\bibinfo {title} {Experimental witnessing for entangled states with limited local measurements},}\ }\href {\doibase 10.1364/PRJ.462212} {\bibfield  {journal} {\bibinfo  {journal} {Photon. Res.}\ }\textbf {\bibinfo {volume} {10}},\ \bibinfo {pages} {2047--2055} (\bibinfo {year} {2022})}\BibitemShut {NoStop}%
\bibitem [{\citenamefont {Cirac}\ and\ \citenamefont {Zoller}(2000)}]{q1}%
  \BibitemOpen
  \bibfield  {author} {\bibinfo {author} {\bibfnamefont {J.~I.}\ \bibnamefont {Cirac}}\ and\ \bibinfo {author} {\bibfnamefont {P.}~\bibnamefont {Zoller}},\ }\bibfield  {title} {\enquote {\bibinfo {title} {A scalable quantum computer with ions in an array of microtraps},}\ }\href {\doibase 10.1038/35007021} {\bibfield  {journal} {\bibinfo  {journal} {Nature}\ }\textbf {\bibinfo {volume} {404}},\ \bibinfo {pages} {579--581} (\bibinfo {year} {2000})},\ \bibinfo {note} {published April 1, 2000}\BibitemShut {NoStop}%
\bibitem [{\citenamefont {Clark}\ \emph {et~al.}(2009)\citenamefont {Clark}, \citenamefont {Lin}, \citenamefont {Brown},\ and\ \citenamefont {Chuang}}]{q2}%
  \BibitemOpen
  \bibfield  {author} {\bibinfo {author} {\bibfnamefont {R.~J.}\ \bibnamefont {Clark}}, \bibinfo {author} {\bibfnamefont {T.}~\bibnamefont {Lin}}, \bibinfo {author} {\bibfnamefont {K.~R.}\ \bibnamefont {Brown}}, \ and\ \bibinfo {author} {\bibfnamefont {I.~L.}\ \bibnamefont {Chuang}},\ }\bibfield  {title} {\enquote {\bibinfo {title} {A two-dimensional lattice ion trap for quantum simulation},}\ }\href {\doibase 10.1063/1.3056227} {\bibfield  {journal} {\bibinfo  {journal} {Journal of Applied Physics}\ }\textbf {\bibinfo {volume} {105}},\ \bibinfo {pages} {013114} (\bibinfo {year} {2009})}\BibitemShut {NoStop}%
\bibitem [{\citenamefont {Mielenz}\ \emph {et~al.}(2016)\citenamefont {Mielenz}, \citenamefont {Kalis}, \citenamefont {Wittemer}, \citenamefont {Hakelberg}, \citenamefont {Warring}, \citenamefont {Schmied}, \citenamefont {Blain}, \citenamefont {Maunz}, \citenamefont {Moehring}, \citenamefont {Leibfried},\ and\ \citenamefont {Schaetz}}]{q3}%
  \BibitemOpen
  \bibfield  {author} {\bibinfo {author} {\bibfnamefont {M.}~\bibnamefont {Mielenz}}, \bibinfo {author} {\bibfnamefont {H.}~\bibnamefont {Kalis}}, \bibinfo {author} {\bibfnamefont {M.}~\bibnamefont {Wittemer}}, \bibinfo {author} {\bibfnamefont {F.}~\bibnamefont {Hakelberg}}, \bibinfo {author} {\bibfnamefont {U.}~\bibnamefont {Warring}}, \bibinfo {author} {\bibfnamefont {R.}~\bibnamefont {Schmied}}, \bibinfo {author} {\bibfnamefont {M.}~\bibnamefont {Blain}}, \bibinfo {author} {\bibfnamefont {P.}~\bibnamefont {Maunz}}, \bibinfo {author} {\bibfnamefont {D.~L.}\ \bibnamefont {Moehring}}, \bibinfo {author} {\bibfnamefont {D.}~\bibnamefont {Leibfried}}, \ and\ \bibinfo {author} {\bibfnamefont {T.}~\bibnamefont {Schaetz}},\ }\bibfield  {title} {\enquote {\bibinfo {title} {Arrays of individually controlled ions suitable for two-dimensional quantum simulations},}\ }\href {\doibase 10.1038/ncomms11839} {\bibfield  {journal} {\bibinfo  {journal} {Nature Communications}\ }\textbf {\bibinfo {volume} {7}},\ \bibinfo
  {pages} {ncomms11839} (\bibinfo {year} {2016})}\BibitemShut {NoStop}%
\bibitem [{\citenamefont {Bruzewicz}\ \emph {et~al.}(2016)\citenamefont {Bruzewicz}, \citenamefont {McConnell}, \citenamefont {Chiaverini},\ and\ \citenamefont {Sage}}]{q4}%
  \BibitemOpen
  \bibfield  {author} {\bibinfo {author} {\bibfnamefont {C.~D.}\ \bibnamefont {Bruzewicz}}, \bibinfo {author} {\bibfnamefont {R.}~\bibnamefont {McConnell}}, \bibinfo {author} {\bibfnamefont {J.}~\bibnamefont {Chiaverini}}, \ and\ \bibinfo {author} {\bibfnamefont {J.~M.}\ \bibnamefont {Sage}},\ }\bibfield  {title} {\enquote {\bibinfo {title} {Scalable loading of a two-dimensional trapped-ion array},}\ }\href {\doibase 10.1038/ncomms13005} {\bibfield  {journal} {\bibinfo  {journal} {Nature Communications}\ }\textbf {\bibinfo {volume} {7}},\ \bibinfo {pages} {13005} (\bibinfo {year} {2016})}\BibitemShut {NoStop}%
\bibitem [{\citenamefont {Hakelberg}\ \emph {et~al.}(2019)\citenamefont {Hakelberg}, \citenamefont {Kiefer}, \citenamefont {Wittemer}, \citenamefont {Warring},\ and\ \citenamefont {Schaetz}}]{q5}%
  \BibitemOpen
  \bibfield  {author} {\bibinfo {author} {\bibfnamefont {F.}~\bibnamefont {Hakelberg}}, \bibinfo {author} {\bibfnamefont {P.}~\bibnamefont {Kiefer}}, \bibinfo {author} {\bibfnamefont {M.}~\bibnamefont {Wittemer}}, \bibinfo {author} {\bibfnamefont {U.}~\bibnamefont {Warring}}, \ and\ \bibinfo {author} {\bibfnamefont {T.}~\bibnamefont {Schaetz}},\ }\bibfield  {title} {\enquote {\bibinfo {title} {Interference in a prototype of a two-dimensional ion trap array quantum simulator},}\ }\href {\doibase 10.1103/PhysRevLett.123.100504} {\bibfield  {journal} {\bibinfo  {journal} {Phys. Rev. Lett.}\ }\textbf {\bibinfo {volume} {123}},\ \bibinfo {pages} {100504} (\bibinfo {year} {2019})}\BibitemShut {NoStop}%
\bibitem [{\citenamefont {Jain}\ \emph {et~al.}(2020)\citenamefont {Jain}, \citenamefont {Alonso}, \citenamefont {Grau},\ and\ \citenamefont {Home}}]{q6}%
  \BibitemOpen
  \bibfield  {author} {\bibinfo {author} {\bibfnamefont {S.}~\bibnamefont {Jain}}, \bibinfo {author} {\bibfnamefont {J.}~\bibnamefont {Alonso}}, \bibinfo {author} {\bibfnamefont {M.}~\bibnamefont {Grau}}, \ and\ \bibinfo {author} {\bibfnamefont {J.~P.}\ \bibnamefont {Home}},\ }\bibfield  {title} {\enquote {\bibinfo {title} {Scalable arrays of micro-penning traps for quantum computing and simulation},}\ }\href {\doibase 10.1103/PhysRevX.10.031027} {\bibfield  {journal} {\bibinfo  {journal} {Phys. Rev. X}\ }\textbf {\bibinfo {volume} {10}},\ \bibinfo {pages} {031027} (\bibinfo {year} {2020})}\BibitemShut {NoStop}%
\bibitem [{\citenamefont {Jain}\ \emph {et~al.}(2024)\citenamefont {Jain}, \citenamefont {Sägesser}, \citenamefont {Hrmo}, \citenamefont {Torkzaban}, \citenamefont {Stadler}, \citenamefont {Oswald}, \citenamefont {Axline}, \citenamefont {Bautista-Salvador}, \citenamefont {Ospelkaus}, \citenamefont {Kienzler},\ and\ \citenamefont {Home}}]{q7}%
  \BibitemOpen
  \bibfield  {author} {\bibinfo {author} {\bibfnamefont {S.}~\bibnamefont {Jain}}, \bibinfo {author} {\bibfnamefont {T.}~\bibnamefont {Sägesser}}, \bibinfo {author} {\bibfnamefont {P.}~\bibnamefont {Hrmo}}, \bibinfo {author} {\bibfnamefont {C.}~\bibnamefont {Torkzaban}}, \bibinfo {author} {\bibfnamefont {M.}~\bibnamefont {Stadler}}, \bibinfo {author} {\bibfnamefont {R.}~\bibnamefont {Oswald}}, \bibinfo {author} {\bibfnamefont {C.}~\bibnamefont {Axline}}, \bibinfo {author} {\bibfnamefont {A.}~\bibnamefont {Bautista-Salvador}}, \bibinfo {author} {\bibfnamefont {C.}~\bibnamefont {Ospelkaus}}, \bibinfo {author} {\bibfnamefont {D.}~\bibnamefont {Kienzler}}, \ and\ \bibinfo {author} {\bibfnamefont {J.}~\bibnamefont {Home}},\ }\bibfield  {title} {\enquote {\bibinfo {title} {Penning micro-trap for quantum computing},}\ }\href {\doibase 10.1038/s41586-024-07111-x} {\bibfield  {journal} {\bibinfo  {journal} {Nature}\ }\textbf {\bibinfo {volume} {627}},\ \bibinfo {pages} {510--514} (\bibinfo {year} {2024})},\ \bibinfo
  {note} {published March 1, 2024}\BibitemShut {NoStop}%
\bibitem [{\citenamefont {Textor}(1978)}]{variance}%
  \BibitemOpen
  \bibfield  {author} {\bibinfo {author} {\bibfnamefont {W.}~\bibnamefont {Textor}},\ }\bibfield  {title} {\enquote {\bibinfo {title} {A theorem on maximum variance},}\ }\href {\doibase 10.1007/BF00673011} {\bibfield  {journal} {\bibinfo  {journal} {nt. J. Theor. Phys.}\ }\textbf {\bibinfo {volume} {17}},\ \bibinfo {pages} {599} (\bibinfo {year} {1978})}\BibitemShut {NoStop}%
\end{thebibliography}%
\end{document}